# Precipitation in lean Mg–Zn–Ca alloys

R. E. Schäublin[1,2,*], M. Becker[1,†], M. Cihova[1], S. S. A. Gerstl[1,2], D. Deiana[3], C. Hébert[4], S. Pogatscher[1,5], P. J. Uggowitzer[1] and J. F. Löffler[1]

[1]Laboratory of Metal Physics and Technology, Department of Materials, ETH Zurich, 8093 Zurich, Switzerland

[2]Scientific Center for Optical and Electron Microscopy, ETH Zurich, 8093 Zurich, Switzerland

[3]Interdisciplinary Center for Electron Microscopy (CIME), École Polytechnique Fédérale de Lausanne, 1015 Lausanne, Switzerland

[4]Laboratoire de Spectroscopie et Microscopie Electronique, Institut de Physique, École Polytechnique Fédérale de Lausanne, 1015 Lausanne, Switzerland

[5]Chair of Nonferrous Metallurgy, Montanuniversitaet Leoben, 8700 Leoben, Austria

[*]Corresponding author.

[†]Deceased.

## Abstract

While lean Mg–Zn–Ca alloys are promising materials for temporary implants, questions remain on the impact of Zn and Ca. In this context, the precipitation in Mg-1.5Zn-0.25Ca (ZX20, in wt.%) was investigated upon linear heating from about room temperature to 400 °C, with a particular focus on the debated ternary precipitate phase. Three exothermic differential scanning calorimetry (DSC) peaks were observed at 125, 250 and 320 °C. The microstructure at the end of these peaks (205, 260 and 350 °C) was probed in a multiscale correlative approach using atom probe tomography (APT) and transmission electron microscopy (TEM). At 205 °C, APT analysis revealed Ca-rich, Zn-rich and Zn–Ca-rich clusters of about 3 nm in size and with a number density of $5.7 \times 10^{23}$ m$^{-3}$. At 260 °C, APT and TEM showed mono-layered Zn–Ca-rich Guinier–Preston (GP) zones of about 8 nm in size and with a number density of $1.3 \times 10^{23}$ m$^{-3}$. At 350 °C, there are larger, highly coherent elongated precipitates of about 50 nm in size, in a lower number density, and of two types: either binary Mg$_2$Ca precipitates or less numerous ternary Ca$_2$Mg$_6$Zn$_3$ precipitates, as deduced from scanning TEM-based energy dispersive X-ray spectrometry (EDS) and nanodiffraction in TEM. The highest hardening, probed by Vickers testing, relates to the end of the 125 °C DSC peak and thus to GP zones, which outperform the hardening induced by the nanoscale clusters and the larger intermetallic particles. Precipitation in ZX20 upon linear heating was compared to the one induced by hot extrusion at 330 °C. Here, ternary precipitates outnumber the binary ones and are larger, incoherent, and in a much lower number density. They are unequivocally made of hexagonal $Ca_2Mg_5Zn_5$, as deduced by atomically resolved EDS mapping and scanning TEM imaging, and supported by simulations. The complexity of precipitation in lean Mg–Zn–Ca alloys and its kinetic path are discussed.

## 1. Introduction

Magnesium (Mg)-based alloys with a density of 1.7-2.0 g·cm$^{-3}$ are the lightest commonly used structural materials, making them interesting for use in transport to increase energy efficiency, and their biodegradation properties make them interesting for temporary implants, for instance for osteosynthesis. Here the leanness of the alloy is highly desired in order to reduce costs and biotoxicity. The applicability of Mg alloys in both fields, light-weight structural applications and biomedicine, is, however, often limited by their relatively low strength. Precipitation hardening is one effective way to enhance the strength of Mg alloys, and as precipitates affect corrosion properties, precipitation may also be used to tailor their degradation rate depending on the desired biomedical applications. A thorough understanding of precipitation and its effect on the material's macroscopic properties is thus essential.

Addition of calcium (Ca) to Mg is known to improve castability, and strength and creep resistance [1], the latter being due to precipitates made of the binary Mg–Ca phase [2, 3]. This phase is well identified: it is the stoichiometric compound $Mg_2Ca$, which has a C14 hexagonal structure, space group $P6_3/mmc$, and lattice parameters $a = 0.62613$ nm and $c = 1.01573$ nm. Corrosion and oxidation resistance is also enhanced, due to the formation of a protective oxide layer, and the flammability of the molten Mg alloy is reduced [4, 5]. Furthermore, the addition of up to 0.3 wt.% Ca enhances ductility through grain refinement occurring during solidification and extrusion [6]. The extent of efficient age hardening is, however, limited because of the coarse nature of the $Mg_2Ca$ precipitates induced by the low solubility of Ca in Mg [2, 3].

The addition of Zn to the Mg–Ca system increases hardness by solid-solution strengthening and by enhancing the precipitation-hardening ability via the formation of precipitates made of the binary Mg–Ca and/or a ternary Ca–Mg–Zn phase [2]. Conversely, the addition of Ca to Mg–Zn alloys enhances their hardening ability by refining the precipitates [7, 8]. Thanks to Zener drag, precipitates in Mg–Zn–Ca alloys restrict grain-boundary movement [9, 10], which permits to maintain small grain sizes during dynamic recrystallization in hot extrusion or during static recovery and recrystallization in thermal treatments [10].

The formation of these secondary-phase precipitates thus results in improved strength and simultaneously enhanced ductility, making Mg–Zn–Ca alloys very interesting in terms of their mechanical properties. In addition, they are promising candidates for biodegradable implant applications because all three elements are essential elements to the human body and are considered in small quantities to be bioresorbable [11-14], with even positive responses in terms of their biocompatibility [15-19]. The mechanical properties of Mg alloys are more favorable than the ones of commonly used metallic biomaterials, such as stainless steel, titanium or cobalt–chromium alloys, because their Young's modulus is similar to that of bone (~45 GPa). It has also been observed that bone mass increases and that mineral apposition rates are high around degrading Mg-bone implants [18, 20]. They are thus considered to be ideal for fracture fixation due to reduced stress-shielding effects and enhanced bone growth [13, 18, 20, 21]. Interestingly, the two phases appearing in lean Mg–Zn–Ca alloys, which are Ca-rich and Zn–Ca rich ones, form precipitates that have an opposite electrode potential relative to the Mg matrix. With the Ca-rich phase being less noble and the Ca−Zn-rich phase being more noble, this respectively reduces or accelerates the corrosion of the Mg matrix by means of galvanic coupling. Thus, an adjustment of the degradation rate may be achieved by producing the appropriate type of precipitation via a suitable heat treatment. To that end, a detailed understanding is required concerning the precipitates structure in order to fully understand the associated corrosion susceptibility of the alloy [18, 22, 23] and the related corrosion mechanisms occurring at the nanometer scale, such as those recently unveiled [24, 25].

## 2. Structure of the ternary Ca–Mg–Zn phase in Mg–Zn–Ca alloys

While the binary Mg–Ca phase is well known, the ternary Ca-Mg–Zn precipitate phase is despite intensive studies still discussed, with structural and compositional characteristics that seem to strongly depend on the alloy's nominal composition, Zn–to–Ca ratio, and processing. The essence of this long-standing debate is outlined in the following. In 1933, Paris [26] presented the first results on the Mg–Zn–Ca system, investigating 189 different alloys. Through metallography following heat treatments [27] he identified one ternary compound in large polygons with a melting point of 495 °C as $Ca_2Mg_5Zn_5$, and henceforth designated it 2-5-5, but did not provide any crystallographic data. Clark [28] presented in 1961 a 335 °C isothermal section of the Mg–Zn–Ca phase diagram studying 76 key alloys. Using X-ray diffraction (XRD) and metallography he observed two ternary phases, namely β ($Ca_2Mg_6Zn_3$) and ω ($Ca_2Mg_5Zn_{13}$), and concluded that they differ from Paris' suggestion. Although no structural data was given, these two phases were incorporated to the Joint Committee on Powder Diffraction Standards (JCPDS) as Card 12-0266 [29].

In the study of Nie and Muddle from 1997 [2], it was shown that precipitation hardening following aging at 200 °C of the binary alloy Mg-1Ca (wt.%) was substantially enhanced by the addition of 1 wt.% Zn. The authors explained this via a refinement of the precipitates' distribution, considering the atomic radii of the elements. With an atomic radius for Ca, Mg and Zn of respectively 0.180, 0.150 and 0.135 nm [30] and with the assumption of a coherent hexagonal structure for the $Mg_2Ca$ phase, the incorporation of large Ca atoms generates a rather large lattice mismatch with the Mg matrix. It was thus suggested that inclusion of small Zn atoms into the $Mg_2Ca$ unit cell, making a ternary phase, reduces this lattice misfit. The improved lattice matching would in turn lead to an enhanced rate for homogeneous nucleation and thus to a finer distribution of precipitates.

Larionova *et al.* [31] investigated several melt-spun Mg–Zn–Ca alloys (1.25-4.25 at.% Zn, 0.83-3.19 at.% Ca) and showed that the XRD reflections of a ternary phase after heat treatment closely match the reference data of the $Ca_2Mg_6Zn_3$ phase from the *JCPDS* Card 12-0266 [29]. Although still no full crystallographic information was obtained, the authors suggested an hexagonal structure. Using energy dispersive X-ray spectrometry (EDS) in scanning electron microscopy (SEM) the composition of that phase was determined to be $Ca_{14}Mg_{56}Zn_{30}$ (~$Ca_2Mg_8Zn_{4.3}$). The slight mismatch observed between its XRD spectrum peaks and the ones of $Ca_2Mg_6Zn_3$ was attributed to a variation of the interplanar distance caused by the slight difference in Zn or Ca content. It was observed that an increase in the Zn content led to a contraction and an increase in Ca to an expansion of the lattice, while preserving the suggested hexagonal structure.

Jardim *et al.* in 2002 [32, 33] analyzed the ternary phase of precipitates in melt-spun Mg-6Zn-1.5Ca. Even though its composition was close to $Ca_2Mg_4Zn_3$ (43.8 ± 3.3 at.% Mg, 23.2 ± 1 at.% Ca and 33 ± 2.4 at.% Zn), they assumed the phase to be $Ca_2Mg_6Zn_3$ and attributed the 20% difference in composition to the incertitude in the standard-less quantitative EDS analysis. From selected area diffraction patterns (SADPs) they concluded that the crystal has a six- or threefold symmetry and therefore the possible crystal structures to be cubic, trigonal or hexagonal. The authors found in the literature a compound with atoms of similar radius and same stoichiometry, namely the compound $Si_2Te_6Mn_3$, which has a trigonal structure and space group P$\bar{3}$1c, and used it as template for the compound Ca$_2$Mg$_6$Zn$_3$. They concluded from a fairly good match between the simulated and experimental diffraction patterns obtained by a lattice parameters adjustment, that the Ca$_2$Mg$_6$Zn$_3$ structure is trigonal with space group P$\bar{3}$1c. It should be noted, however, that the lattice parameters they derived ($a$ = 9.7 Å, $c$ = 10.0 Å) differ strongly from the ones of Si$_2$Te$_6$Mn$_3$ ($a$ = 7.029 Å, $c$ = 14.255 Å) and thus imply a compression along the $c$ axis with a Ca–Ca shortest interatomic distance of 1.6 Å, which may be unrealistic. Nevertheless, the phase identification performed by SADP in the TEM was later confirmed by XRD, for which the measured peaks were close to the reference data from the *JCPDS* Card 12-0266 [29]. Levi *et al.* [34] and Bamberger *et al.* [35] concluded that the isothermal age hardening observed for other Mg–Zn–Ca alloys (Mg-3.2Zn-1.6Ca at 150°C, 175°C, 200°C and 225°C; Mg-1.05Zn-0.5Ca and Mg-1.9Zn-0.3Ca at 175°C) was due to the precipitation of Ca$_2$Mg$_6$Zn$_3$, while the precipitation of Mg$_2$Ca had no or little effect (in Mg-0.2Zn-0.5Ca). In contrast, Oh *et al.* [36] showed by TEM analysis that the age hardening at 200 °C in the under-aged and peak-aged state for an Mg-0.8Zn-0.5Ca alloy can be attributed to metastable, internally ordered, plate-like, monolayered Guinier–Preston (GP) zones on the basal plane. Complementary atom probe tomography (APT) analysis revealed them to consist of a Ca–Mg–Zn phase with an average composition of Mg–18Ca–8Zn (in at.%.) It was concluded that the negative mixing enthalpy of Ca and Zn and the coupled compensation of the lattice misfit were the driving force for the precipitation of this ordered structure. In the over-aged state, the size and thickness of the GP zones increased, nanoscale precipitates disappeared, and the Mg$_2$Ca phase containing Zn occurred in the form of precipitates on the basal plane. Oh-ishi *et al.* [3] also observed by TEM in an Mg-1.6Zn-0.5Ca alloy that the GP zones are the sole contributor to age hardening at 200 °C, but in this alloy the GP zones were only slightly enlarged via over-aging and retained the same structure. The authors concluded that the growth of the GP zones was suppressed by their good match to the matrix lattice, while some coarse precipitates on the basal plane were also observed but not investigated further. In contrast, in the Mg-4.2Zn-0.5Ca alloy no GP zones were found and the comparably marginal age-hardening response was attributed to coarse Ca$_2$Mg$_6$Zn$_3$ precipitates. The authors thus concluded that the formation of ordered GP zones is suppressed in high-Zn-containing alloys due to the formation of Ca$_2$Mg$_6$Zn$_3$ [3].

Because the reported crystal structure and composition range of the ternary compound have been contradictory [3, 27-29, 31-33], Zhang et al. [37] investigated it again via the diffusion-couple method and concluded that the compound is not stoichiometric but exhibits a composition range, i.e. Ca$_3$Mg$_x$Zn$_{15-x}$ (4.6 ≤ $x$ ≤ 12), following its heat treatment at 335 °C. In this compound, named IM1, the lattice parameters increase linearly with increasing Mg content. The prototype structure taken for IM1 was $Sc_3Ni_{11}Si_4$, which has a hexagonal $P6_3/mmc$ structure (space group 194). In another work of Zhang *et al.* [38], three additional ternary intermetallic compounds were identified, namely $Ca_{14.5}Mg_{15.8}Zn_{69.7}$ (IM2), $Ca_2Mg_5Zn_{13}$ (IM3) and $Ca_{1.5}Mg_{55.3}Zn_{43.2}$ (IM4). Here, IM2 and IM4 were considered to be stoichiometric compounds, while IM3 showed extended ternary solubility with a composition range of 8.2-9.1 at.% Ca, 27.1-31.0 at.% Mg and 60.8-64.7 at.% Zn (both Ca and Mg atoms were substituted by Zn atoms). Furthermore, it was concluded that the binary compounds $CaZn_{11}$ and $CaZn_{13}$ can be extended to ternary phases via Mg incorporation, with a maximum solid Mg solubility of 8.4 and 15.5 at.%, respectively [38]. In the $Mg_2Ca$ $C_{14}$ Laves phase both Ca and Mg atoms may be substituted by Zn to its maximum ternary solubility, resulting in $Ca_{33.3}Mg_{55.9}Zn_{10.8}$ [39]. This is, however, contradicted in [38], where enrichment of Zn in $Mg_2Ca$ is assumed to proceed by the replacement of Mg atoms only.

Li *et al.* [40] investigated equilibrium phases in the Mg–Zn–Ca system at 300 °C with Zn between 10 and 62 at.% and Ca between 5 and 15 at.%. They found two ternary phases, $T_1$ and $T_2$, in equilibrium with the Mg-based solid solution, both having a hexagonal structure. The phase $T_1$ has a Zn content between 20.5 and 48.9 at.% and Ca stays nearly constant (13.9 – 15.2 at.%), while $T_2$ has a narrow composition range (63.2 – 65.6 at.% Zn, 7.1 – 8.4 at.% Ca, Mg for the balance). $T_1$ appeared in all alloys with 40 at.% Zn or less, while $T_2$ was only found in alloys containing at least 40 at.% Zn.

Langelier *et al.* in 2012 [41] analyzed the alloy Mg-0.9Zn-2.1Ca during non-isothermal aging. They concluded that its age-hardening potential in the peak-aged condition is due to fine GP zones. However, these were not resolved with the TEM used and only some weak background signal in the SADP hinted at them. The authors reported that with further aging the GP zones coarsened first to fine basal plates in co-presence with unidentified nanoscale precipitates, and then to large blocky precipitates and large basal plates made of $Mg_2Ca$. With further aging, the nanoscale precipitates disappeared. Although the ternary $Ca_2Mg_6Zn_3$ phase appeared in their thermodynamic calculations using FactSage [42] with the database of their time, no evidence for it was found in their work.

Kubok *et al.* [43] investigated the composition range of ternary precipitates in an Mg-3Zn-*x*Ca alloy, with *x* from 0 to 1.3. In the ternary phase they found an increase of the Zn-to-Ca ratio from 1.5 for Mg-3Zn-1.3Ca to 3.84 for Mg-3Zn-0.2Ca, while the Ca content remained at about the same level (14.1 to 15.2 at.%). This corresponds to an increase of Zn from 22.8 to 54.1 at.% for a slightly increasing Ca content. It was suggested that the addition of Ca to the Mg–Zn system changes the solubility of Zn in Mg and increases the lattice parameters.

Recently, we attempted to clarify the ternary phase with an alloy made with a nominal composition of 12.4 at.% Ca, 58.15 at.% Mg and 29.45 at.% Zn [44], which falls between the $Ca_2Mg_6Zn_3$, $Ca_2Mg_5Zn_5$ and IM1 phase fields according to the respective Mg–Zn–Ca phase diagrams published by Paris [27], Clark [28] and Zhang [37]. By structure refinement based on XRD, and assisted by EDS and electron diffraction in a TEM, we identified that the main resulting compound is $Ca_2Mg_5Zn_5$ with a hexagonal structure $P6_3/mmc$ (space group 194) [44], and not trigonal $Ca_2Mg_6Zn_3$. However, the phase was in bulk form and not in dispersed precipitates as found in lean Mg–Zn–Ca alloys. There is thus still an uncertainty on the nature of the ternary phase precipitates, despite the many efforts devoted to it (see e.g. [45]). Table 1 summarizes the candidate phases that, to the best of our knowledge, have been postulated so far. Moreover, a clear understanding of the pathway leading to these precipitates has been still lacking.

| Compound | Composition | Crystal | Prototype | Reference | |
|---|---|---|---|---|---|
| $Ca_2Mg_5Zn_5$ | $Ca_2Mg_5Zn_5$ | - | - | Paris 1933 | [25, 26] |
| $Ca_2Mg_6Zn_3$ | $Ca_2Mg_6Zn_3$ | - | - | Clark 1961 | [27] |
| $Ca_2Mg_5Zn_{13}$ | $Ca_2Mg_5Zn_{13}$ | - | - | Clark 1961 | [27] |
| CaMgZn | CaMgZn | Hexagonal | - | Schulze 1961 | [63] |
| $Ca_2Mg_6Zn_3$ | $Ca_2Mg_6Zn_3$ | Trigonal | $Si_2Te_6Mn_3$ | Jardim 2002 | [31] |
| IM1 | $Ca_3Mg_xZn_{15-x}$ (4.6 ⩽ x ⩽ 12 at 335°C) | - | - | Zhang 2010 | [36] |
| | $Ca_3Mg_{11}Zn_4$ | Hexagonal | $Sc_3Ni_{11}Si_4$ | | |
| | $Ca_2Mg_6Zn_3$ | Trigonal | $Si_2Te_6Mn_3$ | | |
| | $CaMgZn_3$ | Hexagonal | $YMgZn_3$ | | |
| IM2 | $Ca_{14.5}Mg_{15.8}Zn_{69.7}$ | | - | Zhang 2010, 2011 | [36, 37] |
| IM3 | $Ca_2Mg_5Zn_{13}$ | Hexagonal | - | Zhang 2011 | [37] |
| IM4 | $Ca_{1.5}Mg_{55.3}Zn_{43.2}$ | | - | Zhang 2011 | [37] |
| $Ca_2Mg_5Zn_5$ | $Ca_{16}Mg_{42}Zn_{42}$ | Hexagonal | Refined | Cao 2016 | [43] |

Table 1. Main phases in the Ca−Mg−Zn ternary system considered for the ternary precipitates in Mg–Zn–Ca-lean alloys.

The aim of this work is thus to describe the precipitation sequence in a lean Mg–Zn–Ca alloy with a particular focus on the structure of the ternary Ca–Mg–Zn precipitates, using differential scanning calorimetry (DSC) complemented by APT and TEM. The identified precipitates were further considered to interpret the results of Vickers-hardness tests used to assess the mechanical properties. For this study Mg-1.5Zn-0.25Ca (in wt.%) referred to as 'ZX20' was chosen as a representative for ZX-lean alloys, which shows both adequate mechanical properties and suitable biocorrosion properties in *in vitro* and *in vivo* conditions in the as-extruded state [46]. The composition and structure of the precipitates were assessed and are discussed with respect to existing thermodynamic calculations and possible precipitation kinetics paths, and by comparing the heat-treated to the extruded material. The size and number density of the precipitates are used to establish a relationship between hardness and microstructure.

## 2. Experimental

ZX20 (Mg-Zn1.5(0.56)-Ca0.25(0.15) (in wt.% (at.%))) was prepared at the AIT Austrian Institute of Technology, Light Metals Technologies Ranshofen, Austria, by alloying high-purity elements Zn (99.9999%) and Ca (99.99%) to distilled ultrahigh-purity Mg (99.999%) [47] and subsequent hot extrusion. Prior to extrusion, the billet was homogenized and solutionized at 350 °C for 12 h and 450 °C for 8 h, respectively, followed by a heat treatment at 250 °C for 30 minutes. The billet was indirectly extruded at 330 °C at a ram speed of 18 mm min$^{-1}$ into rods of 6 mm in diameter. The rods were cut into bars of about 20 mm length and subjected to a solution-heat treatment at 500 °C for 2 h in Ar-filled quartz tubes followed by water quenching. They were then stored in LN$_2$ until further use. DSC, Vickers-hardness test, TEM and APT samples were extracted from these bars by spark erosion. The samples were then individually shaped, heat-treated and analyzed as explained in the following.

The DSC measurements were performed with a Mettler-Toledo DSC 1 calorimeter under Ar flow. The samples were in the form of 6 mm disks, about 0.5 mm thick and with a weight of about 25 mg. The reference cell was filled with the same ultrahigh-purity Mg as used in the alloy and with the same shape. The heat-flow curve was obtained by subtracting the curve from a second DSC run from the first measurement. The measurements were performed from 5 °C to 400 °C at a rate of 20 K min$^{-1}$. DSC revealed three exothermic peaks, namely at 125, 250 and 320 °C, finishing at respectively 205, 260 and 350 °C. Additional DSC measurements were obtained with a Netzsch calorimeter to obtain the solidus and melting temperatures.

Vickers hardness ($H_v$) was measured on samples with a diameter of 6 mm and thickness of 0.5 mm from the solution-heat-treated state to various heat-treated ones. They were heated in the Mettler-Toledo DSC at 20 K min$^{-1}$ to 25 °C, 50 °C and then up to 350 °C in 30 K steps, each followed by a water quench prior to the measurement. The hardness of each sample was measured at 10 separate indentations with a 300 g load for a loading time of 6 s with a Gnehm Brickers 220 device. Prior to the measurement, the samples were gently polished with SiC paper to remove contamination and oxides, and rinsed with ethanol.

The 3 mm disks cut for TEM by spark erosion were thinned down by grinding with SiC paper to a thickness of about 100 µm (weight of about 1.2 mg). They were heated in the Mettler-Toledo DSC at 20 K min$^{-1}$ to temperatures corresponding to the end of the DSC peaks (205, 260 and 350 °C) and quenched in water. Ion milling was then used to reach electron transparency deploying a Gatan® Precision Ion Polishing System II (PIPS II). It was performed at LN$_2$ temperature and at 4 kV at an incidence angle of 3.5° until a hole appeared, followed by 20 min at 500 V for final polishing.

TEM analyses were conducted with a Philips Tecnai F30 FEG operated at 300 kV (ScopeM, ETHZ) for high-resolution (HR) TEM using phase contrast. Bright-field (BF) imaging and high-throughput chemical mapping applying large-collection-angle X-ray energy dispersive spectrometry ('Super X' EDS) were performed on a FEI Tecnai Osiris operated at 200 kV (CIME, EPFL) and on a FEI Talos F200A operated at 80 kV (ScopeM, ETHZ). HR scanning TEM (HR-STEM) imaging and corresponding STEM-EDS chemical mapping were carried out on an FEI Themis operated at 80 kV (CIME, EPFL). Structure analysis using SADPs or nanodiffraction was performed on the FEI Talos. The operating voltage was reduced when possible to 80 kV in order to avoid radiation damage, because the threshold for atomic displacement in Mg alloys is 100 kV [48]. Diffraction patterns were interpreted using the electron diffraction analysis software CrysTBox [49] as well as jEMS V4 [50]. The number density of the precipitates was obtained by manually counting them in TEM or STEM micrographs, thus obtaining an area-number density, and by measuring the thickness of the measured region using electron energy loss spectroscopy (EELS) on the F30 or by stereography on the Talos instrument. The volume-number density of the particles is then the area-number density divided by the thickness. The size of the precipitates was measured manually in the TEM or STEM micrographs using ImageJ software.

For APT, matchsticks of 15×0.3×0.3 mm$^3$ cut by spark erosion were slightly ground with 4000 SiC paper (weight of 2.3 mg each). They were heated in the Mettler-Toledo DSC at 20 K min$^{-1}$ to temperatures corresponding to the end of the DSC peaks (205, 260 and 350 °C) and quenched in water. They were then electrochemically polished in a standard two-step procedure to obtain a needle-like specimen shape, first in a solution of 10 ml perchloric acid (60 vol.%) in 90 ml methanol, and then sharpened in a solution of 1 ml perchloric acid in 110 ml 2-butoxyethanol at room temperature and at a DC voltage of 20 V.

The APT analyses were conducted with a LEAP 4000X HR (Cameca®) instrument at a specimen temperature of 64 K under ultrahigh vacuum ($1.5 \times 10^{-11}$ mbar) in voltage-pulsing mode, applying a pulse fraction of 20%. Data was reconstructed and analyzed with an IVAS 3.6.12 software package (Cameca®). Detailed analyses were performed on sub-volumes, avoiding known crystallographically induced artifacts. Clusters and precipitates were then identified by applying a combination of cluster-search algorithms based on nearest-neighbor (NN) distributions, and via isoconcentration surfaces. The composition of these features were determined via counting statistics and the proximity-histogram (proxigram) method [51] applying a 0.5 nm bin size. The cluster-search algorithm is based on the maximum separation method [52], applying the following parameters following the procedure suggested by Vaumousse *et al.* [53] and iterative refinement steps in IVAS: $d_{max}$ = 1.4 nm (maximum spacing between atoms making a cluster), $n$ = 1 (order parameter), $N_{min}$ =30 (minimum number of atoms defining a cluster), $L$=1.4 nm (envelope distance), $d_{erosion}$ =0.7 nm (erosion distance), with 1 nm voxel size and 5 nm delocalization.

## 3. Results

Figure 1 shows the calculated fraction of the phases occurring in ZX20 as a function of temperature. For that purpose, Pandat™ software was used, where with the input from the MatCalc database mc_mg_v1.0009 the IM1 (as $Ca_2Mg_6Zn_3$), IM3 (as $Ca_2Mg_5Zn_{13}$), $Mg_2Ca$ and the liquidus phase are considered. At low temperatures only the ternary phases are present, with the IM3 phase dominating below 130 °C and IM1 dominating above. Beyond about 230 °C IM1 is the only phase. Its volume fraction then decreases, indicating its destabilization with increasing temperature, and vanishes completely above about 380 °C. The $Mg_2Ca$ phase appears at around 360 °C, peaks at about 380 °C and disappears above 400 °C. The calculated solidus temperature is 460 °C. In such calculations, thermodynamic equilibrium is assumed.

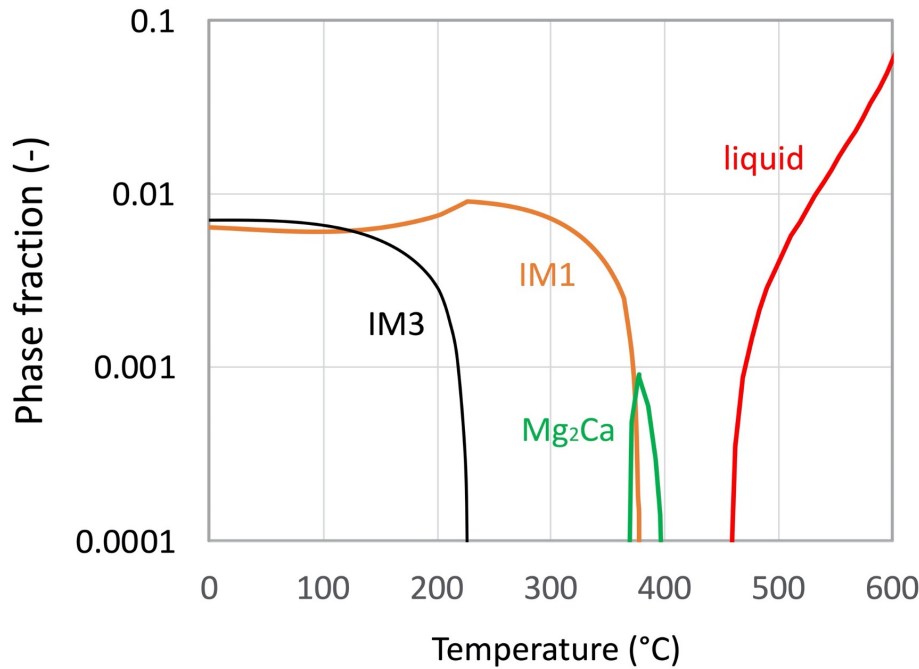

Figure 1. Calculated fraction of phases occurring in ZX20 as a function of temperature. Simulated using the software package Pandat™ with MatCalc™ database mc_mg_v1.0009, which considers exclusively the ternary phases IM1 (as $Ca_2Mg_6Zn_3$) and IM3 (as $Ca_2Mg_5Zn_{13}$), the binary phase $Mg_2Ca$ and the liquid phase.

Figure 2 presents the specific heat flow obtained by DSC on ZX20 as a function of temperature and the related evolution of Vickers hardness ($H_v$). The DSC trace exhibits three main exothermic peaks, namely a first broad and low peak at 125 °C, a second peak at 250 °C and a third peak at 320 °C. Correspondingly, the hardness curve shows a weak increase from about 39 $H_V$ measured at room temperature (corresponding to the solution-heat-treated condition) to 44 $H_v$ at 200 °C, followed by a strong increase reaching the peak hardness of 48 $H_v$ at 290 °C. With a further increase in temperature, the hardness decreases sharply down to 42 $H_v$ at the maximum measurement temperature of 350°C. The error is given by the standard deviation. In order to understand the origin of the three exothermic peaks observed in DSC and the origin of the hardening, the microstructure corresponding to the temperatures at the end of the DSC peaks was investigated by APT and TEM. These temperatures are respectively 205 °C, 260 °C and 350 °C.

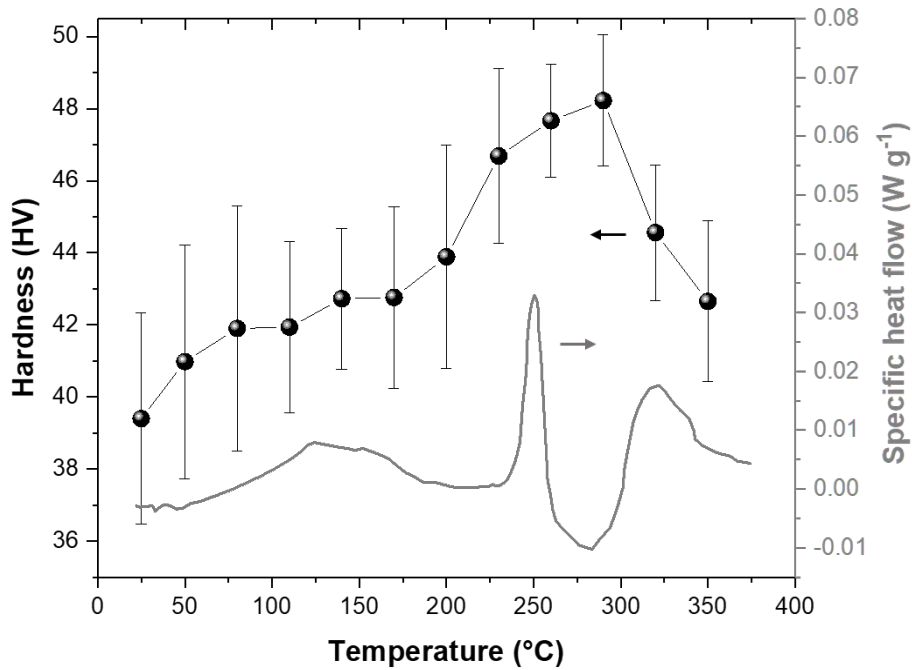

Figure 2. Vickers hardness ($H_V$) and specific heat flow determined by differential scanning calorimetry of ZX20 as a function of temperature at a heating rate of 20 K min$^{-1}$.

Figure 3 presents the microstructural details obtained by APT after heating ZX20 to 205 °C. Note that TEM reveals a microstructure with few dislocations but no precipitates (not shown). In contrast, in the APT reconstruction of the sample, independent clusters of Zn, Ca or Zn+Ca, shown in respectively red, green and yellow, are recognizable when the iso-concentration surface values are set to 0.8% for Zn, 0.4% for Ca and 1.06% for Zn+Ca (all in at.%, Fig. 3a). The atom-map data obtained was further analyzed by considering the pair correlation function to atomic inter-distances up to the fifth-nearest neighbors (5NN) for Zn, Ca and (Zn, Ca) atoms. Fig. 3b shows the 5NN distribution for the (Zn, Ca) atoms of the measured sample versus the distribution in a randomized system based on the same set of atoms. The latter was obtained by distributing the solute atoms stochastically with respect to the fixed positions in the experimental data. A broadening of the distribution and a slight shift of its peak to smaller distances can be observed for the measured data, indicative of clustering. The same observations were made for the distribution of Ca and Zn atoms individually (not shown). This confirms the existence of clusters visible in Fig. 3a. With respect to the cluster types' number density and size, reported in Table 2, the Zn–Ca clusters constitute the majority (~48%) with a size of 3.0 ± 1.4 nm. A large fraction (~30%) of the clusters is Ca-rich with a size of 2.6 ± 0.9 nm, and the rest is Zn-rich (~22%) with a size of 2.8 ± 1.2 nm. The total number density of the clusters is 5.7(± 0.3) × 10$^{23}$ m$^{-3}$.

| T [°C] | Precipitate type | Size [nm] | | Density [m$^{-3}$] | Zn-to-Ca ratio | Method |
|---|---|---|---|---|---|---|
| 205 | globular, Zn-rich | 2.8 ± 1.2 | | 1.3 (± 0.1) × 10$^{23}$ | - | APT |
| | globular, Ca-rich | 2.6 ± 0.9 | | 1.7 (± 0.1) × 10$^{23}$ | - | |
| | globular, Zn−Ca-rich | 3.0 ± 1.4 | | 2.7 (± 0.1) × 10$^{23}$ | 2 | |
| 260 | GP zones, diameter | 7.8 ± 3.9 | | 1.3 (± 0.1) × 10$^{23}$ | 1.2 ± 0.8 (3.3 ± 3.0) | APT (TEM) |
| | | diameter | length | | | |
| 350 | Zn−Ca-rich, small | 9.7 ± 2.6 | 17.0 ± 3.7 | 2.8 (± 0.4) × 10$^{19}$ | 1.4 ± 0.4 | TEM |
| | Ca-rich, small | 12.4 ± 6.2 | 17.8 ± 4.4 | 2.3 (± 0.8) × 10$^{20}$ | 0.6 ± 0.1 | |
| | Zn−Ca-rich, large | 30.5 ± 10.4 | 98 ± 54 | 1.7 (± 1.1) × 10$^{19}$ | 1.5 ± 0.3 | |
| | Ca-rich, large | 25 ± 12 | 92 ± 37 | 1.4 (± 0.4) × 10$^{19}$ | 0.20 ± 0.14 | |

Table 2. Size, number density and Zn-to-Ca atomic-composition ratio of the clusters and precipitates observed in ZX20 heat-treated from solid-solution condition to 350 °C at 20 K min$^{-1}$. At 205 °C, globular nanometric clusters were identified by APT; they were either Zn- or Ca-rich, or Zn–Ca-rich. At 260 °C, atomically thin Zn−Ca-rich Guinier-Preston (GP) zones were observed in TEM and measured in APT for their composition and number density. At 350 °C, elongated Ca-rich or Zn−Ca-rich precipitates were assessed by TEM and EDS.

Fig. 3c shows the proxigrams for the Ca-rich (top graph) and the Zn–Ca-rich (bottom graph) clusters. It appears that for the Ca-rich clusters the content in Ca reaches values much above the one measured in the matrix (0.14 at.%). As for the Zn content in these Ca-defined clusters, a slight increase (~1 at.%) compared to the one measured in the matrix (0.47 at.%) is determined. Conversely, for the Zn–Ca-rich clusters the proxigram shows an enrichment in Zn content (~2 at.%), which dominates the Ca content (~1 at.%).

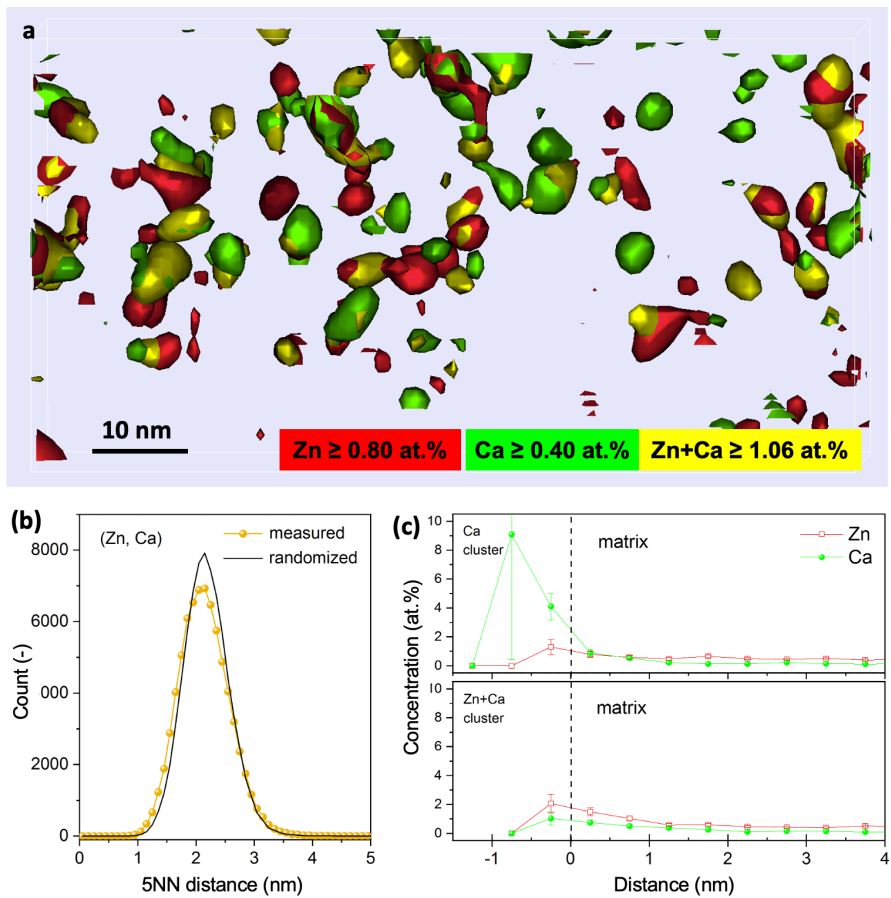

Figure 3. ZX20 heat-treated to 205 °C. (a) APT 3D reconstruction of isoconcentration surfaces for Zn (red), Ca (green) and Zn+Ca (yellow), revealing respective clusters. (b) Fifth-nearest-neighbor (5NN) distribution as a function of distance between Zn and Ca atoms, measured and randomized. (c) Average proxigram of the Zn and Ca concentration in clusters of (top) Ca and (bottom) Zn+Ca.

Figure 4 shows the microstructure characteristic of the material heated to 260 °C. TEM at moderate magnification (not shown) presents a microstructure with few dislocations and no visible precipitates, comparable to the material heat-treated to 205 °C. At higher magnification, however, HAADF STEM imaging along $[2\bar{1}\bar{1}0]$ reveals a fine dispersion of nanometric platelets visible as bright needles. Their bright contrast relates to elements heavier than the Mg matrix. The chemical mapping performed in STEM mode using EDS confirms that they are rich in Ca and Zn (Fig. 4b and c, respectively). BF TEM reveals a diffraction contrast around each of the precipitates (Fig. 4d), indicative of strain in the lattice. High-resolution TEM shows that the precipitates are in fact mainly made of one atomic plane (lower inset in Fig. 4d), and the diffraction pattern obtained (upper inset in Fig. 4d) indicates that their projection is perpendicular to the diffraction vector (0002). This shows that they are Guinier–Preston (GP) zones lying on the basal planes (0001) of the Mg matrix.

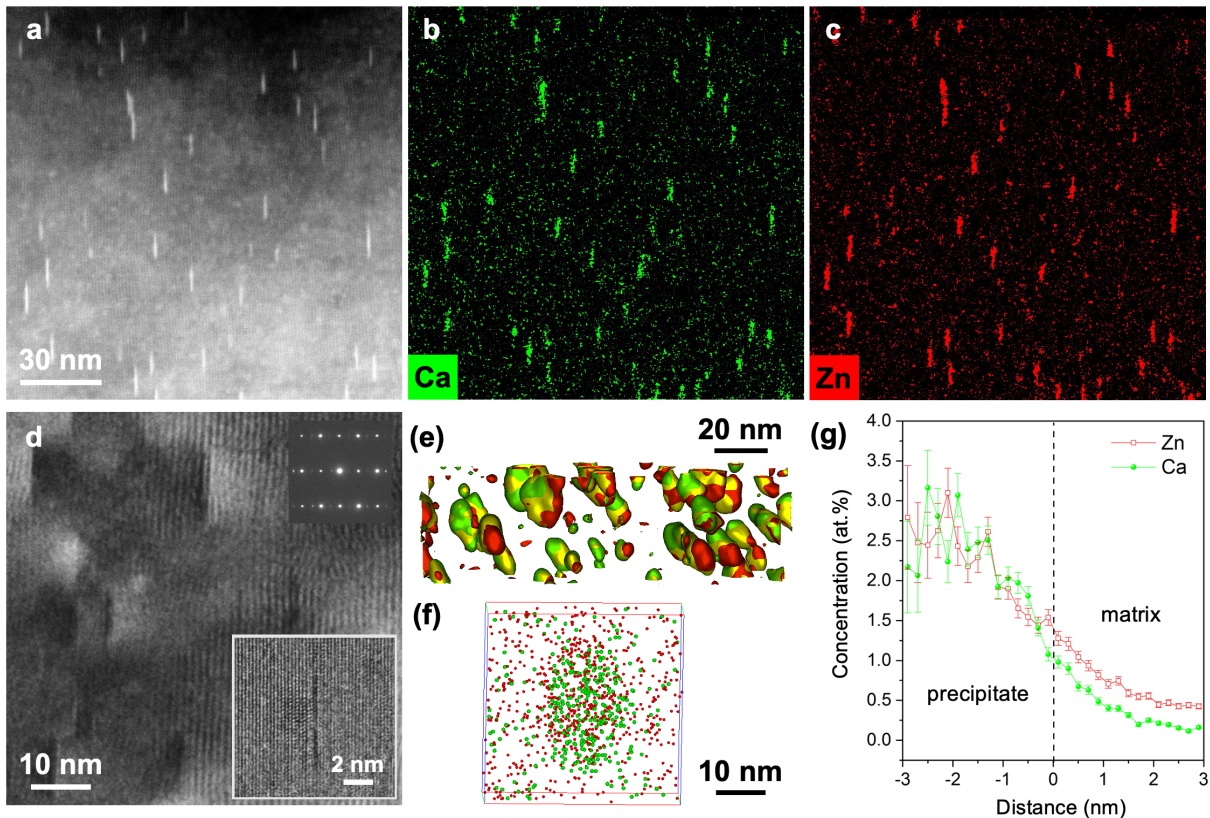

Figure 4. ZX20 heat-treated to 260 °C. (a) HAADF STEM image with (b, c) the corresponding chemical maps revealing the distribution of (b) Ca and (c) Zn. (d) HRTEM image with the corresponding selected-area diffraction pattern in the top right, indicating a beam direction of [$2\bar{1}\bar{1}0$]. The lower inset to (d) shows one precipitate at higher magnification, revealing that the precipitates are monolayered. (e) APT 3D reconstruction of isoconcentration surfaces for (red) Zn at 0.80 at.%, (green) Ca at 0.40 at.%, and (yellow) Zn+Ca at 1.06 at.%. (f) APT zoomed-in view with Zn and Ca atoms, revealing one of the precipitates. (g) Average proxigram of Zn and Ca concentrations in Zn+Ca precipitates.

Fig. 4e shows an APT 3D reconstruction of this sample with isosurfaces set to 0.80 at.% for Zn, 0.40 at.% for Ca and 1.06 at.% for Zn+Ca. Note that the same values were applied on the sample heat-treated to 205 °C (Fig. 3a). The APT reconstruction reveals precipitates that contain both Zn and Ca. With these values 19 clusters are identified, corresponding to a number density of about $6 \times 10^{22}$ m$^{-3}$. Fig. 4f shows the individual Zn and Ca atoms in a volume of 30 nm × 30 nm × 8 nm, isolating one of the precipitates and illustrating that they are Zn- and Ca-rich. In Fig. 4e all precipitates carry simultaneously the signature of Zn, Ca and Zn+Ca. This reveals that all precipitates are made of both Zn and Ca, making them clearly distinguishable from the ones observed in the sample treated at 205 °C (Fig. 3). An additional individual clustering of either Zn or Ca is always seen attached to the Zn–Ca-rich clusters. Here the clustered Zn atoms reside in isoconcentration volumes that are slightly shifted in position relative to the Ca-isoconcentration volumes, with a shift that has the same direction for all precipitates, indicating that in fact they stem from the same Zn–Ca-rich precipitate. This may be due to the locally different evaporation fields of the two elements leading to locally preferential evaporation or retention. All precipitates exhibit an ellipsoidal shape and all have the same orientation, which is consistent with the STEM and TEM observation (Figs. 4a and 4d, respectively). Their length is between 4 and 16 nm and their width between 3 and 8 nm, which correlates well with EDS STEM results, but one should note again that HRTEM revealed that they are GP zones one atomic layer thick (Fig. 4d, inset). This indicates that both STEM and APT render the precipitates much wider than they actually are.

To gain more information about the precipitates the cluster-search algorithm was applied. The analyzed volume was 50 nm × 50 nm × 150 nm. With these parameters the average precipitate contained 253 solute ions and a composition of 96.69 at.% Mg, 1.78 at.% Zn and 1.53 at.% Ca but with large variations, resulting in a Zn-to-Ca ratio of 1.2 ± 0.8. Detailed results are given in Table 2. The total volume fraction of these precipitates was determined to be 0.52%. The threshold value of the isoconcentration surface for Zn + Ca was set to 1.06 at.%. The isoconcentration profile is given in Fig. 4g. It shows an increase of the Zn and Ca concentration from the surface towards the center of an isoconcentration volume. The composition of the matrix can be evaluated with the stabilized values in the proxigram outside of the isoconcentration surfaces. The matrix contains 0.36 ± 0.03 at.% Zn and 0.11 ± 0.03 at.% Ca, with the errors corresponding to the standard deviation.

After heating ZX20 to 350 °C, TEM (Fig. 5) reveals already at low magnification a relatively homogeneous dispersion of precipitates, which are larger in size than those found at 260 °C. They are surrounded by diffraction contrast (Fig. 5a and b), which is either strong (inclined solid white arrow) or weak if any (empty arrow), indicative of some degrees of strain in the surrounding lattice induced by the precipitates. The strain appears in places to be large enough to induce dislocation loop punching, as observed in Fig. 5a (vertical white arrow). The chemical mapping for Ca and Zn (Figs. 5c and d) of the area shown in Fig. 5b reveals that the precipitates are either Ca-rich, making with Mg a binary phase, or Zn–Ca rich, making with Mg a ternary phase. Furthermore, when comparing with the BF TEM micrograph (Fig. 5b), it appears that the binary precipitates are those inducing a strong diffraction contrast around them while it is weak or non-existent for the ternary ones. Note that there are few ternary precipitates that are found attached to dislocation lines; they are larger than the average ones, some reaching about 200 nm (Fig. 5a).

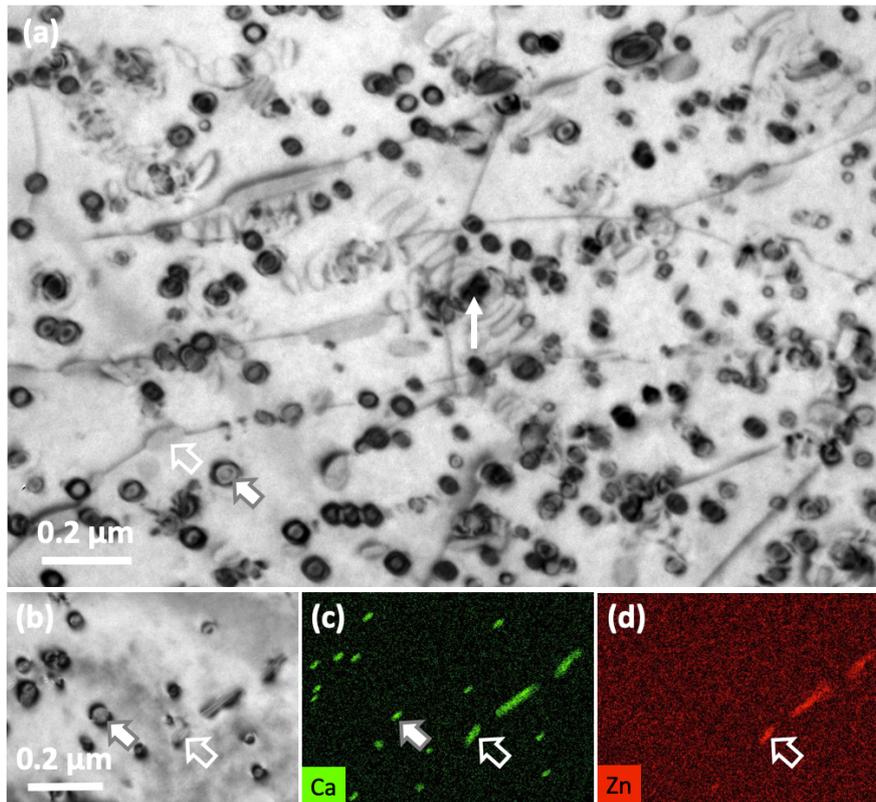

Figure 5. ZX20 heat-treated to 350 °C. (a), (b) BF TEM images revealing a relatively homogeneous dispersion of precipitates, which are surrounded by either a strong diffraction contrast (inclined white arrow) or a weak one, if any (inclined empty arrow). Dislocation loop punching is observed around some of the precipitates ((a), vertical white arrow). (c, d) STEM-EDS chemical maps for (c) Ca and (d) Zn corresponding to the area shown in (b).

The origin of the difference in diffraction contrast between the two types of precipitate can be found at higher magnification. Figure 6 shows typical high-resolution TEM micrographs of a strong-contrast inducing precipitate (binary, Fig. 6a) and a weak-contrast inducing one (ternary, Fig. 6b). Both precipitates seem to be fully coherent with the matrix. However, for the binary precipitate it appears that in the matrix at the precipitate's distal ends there are strong lattice distortions, indicative of the presence of dislocation lines. In contrast, the ternary precipitate seems to blend in the matrix without inducing much visible matrix distortion. However, when the micrographs are viewed along the long axis of the precipitates at an inclined angle, the surrounding matrix lattice is either strongly compressed in the case of the binary precipitate (Fig. 6c) or slightly expanded in the case of the ternary precipitate (Fig. 6d), which relates well to the diffraction contrast observed in BF TEM, which is respectively either strong or weak.

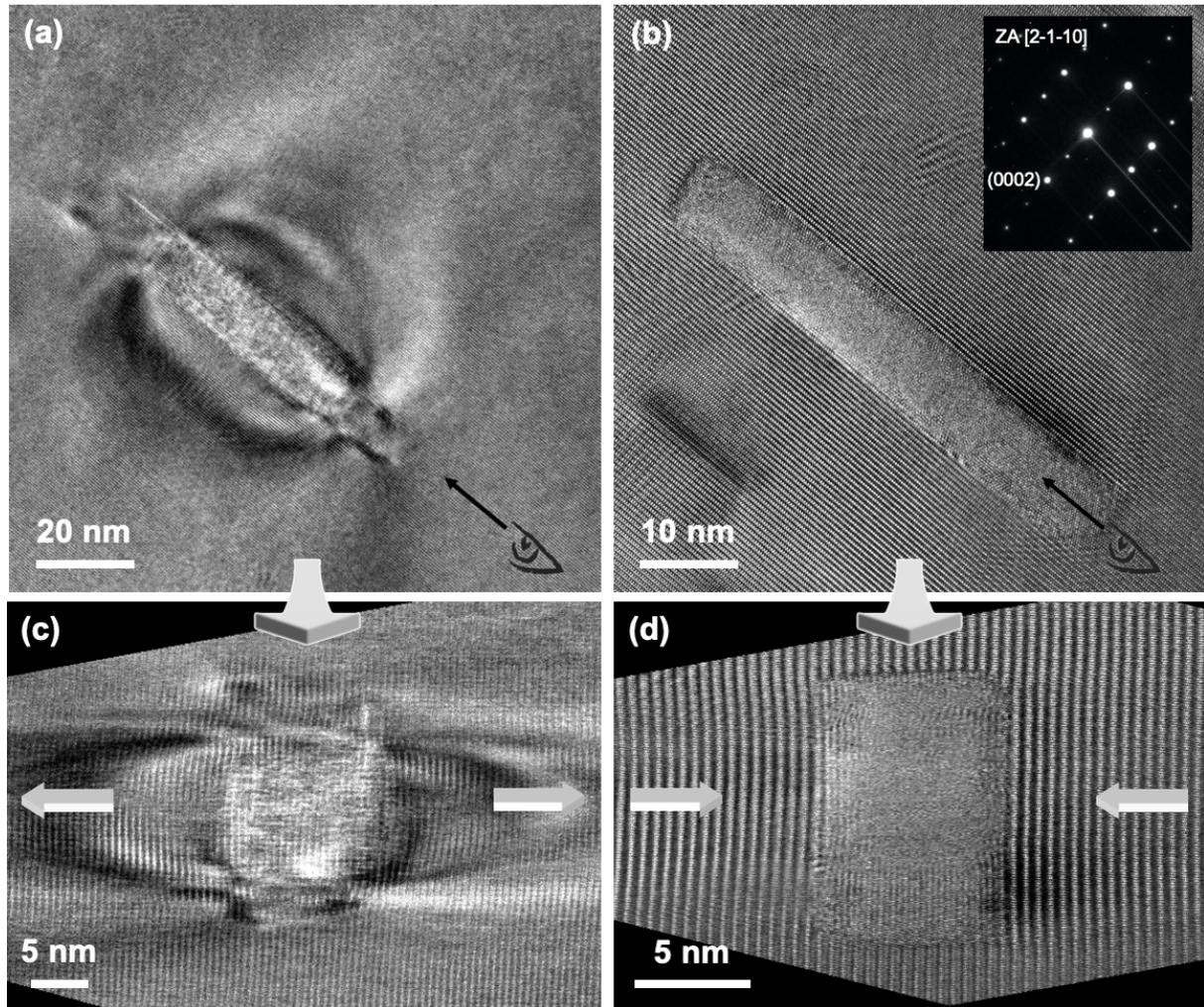

Figure 6. ZX20 heat-treated to 350 °C. (a, b) Typical HRTEM micrograph of a precipitate surrounded by either (a) a strong diffraction contrast or (b) a weak one, if any. Images taken along the zone axis $[2\bar{1}\bar{1}0]$ of the Mg matrix (see diffraction pattern in the inset to (b)). (c, d) Inclined view along the precipitate's long axis (indicated by the eye symbol in (a) and (b)). The images reveal in (c) a compression and (d) an expansion of the matrix around the precipitates, as indicated by the horizontal arrows.

Both types of precipitates, the binary and ternary ones, appear in various sizes, which can be roughly classified as 'small' or 'large' (Table 2). The small precipitates, in average 10 – 12 nm wide and 17 – 18 nm long, have a number density of about $2.3 \times 10^{20}$ m$^{-3}$ for the binary ones and about $2.8 \times 10^{19}$ m$^{-3}$ for the ternary ones. The dimensions of the larger binary precipitates are in average 25 nm in width and 92 nm in length and their number density is around $1.4 \times 10^{19}$ m$^{-3}$. The size of the larger ternary precipitates is in average 30 nm wide and 98 nm in length with a number density of $1.7 \times 10^{19}$ m$^{-3}$, depending on the analyzed region. In sum, while the smaller precipitates seem to have the same shape and size for both types, the larger ones, rectangular in shape, tend to be slightly larger when they are ternary. In all cases the precipitates' EDS indicates that the content in Mg is larger than the one in Zn or Ca. The Zn-to-Ca ratio (Table 2) is in the case of the binary precipitates about 0.6 for the smaller ones and about 0.2 for the larger ones. In the case of the ternary precipitates the Zn-to-Ca ratio amounts to 1.4 for the smaller precipitates and 1.5 for the larger ones. Noting again that we assumed the binary precipitates to be made of Mg$_2$Ca following [2], it appears that for the small ones the contained Zn, giving a Zn-to-Ca ratio of 0.6, is higher than the maximum solubility of Zn in $Mg_2Ca$ of 10.8 at.% found in the literature [39] that gives a ratio of 0.324. This will be elaborated in the discussion.

In the case of the ternary precipitate type, drawing a conclusion on its crystal structure based on literature data was not possible, challenged by the existing debate presented in the introduction. We thus focused on the analysis of the crystal structure of this ternary phase in ZX20. It was analyzed using nanodiffraction and via comparing the obtained diffraction patterns with the ones simulated for all candidates listed in Table 1. The composition derived from the EDS measurements was used to narrow down the candidate structures. Following this protocol, only $Ca_2Mg_6Zn_3$, $Ca_2Mg_5Zn_5$ and $Ca_3Mg_{11}Zn_4$ structures (Fig. S1 in the Supplementary Information) were considered in the following.

Figure 7 presents two nanodiffraction patterns experimentally obtained in two different ternary precipitates for two zone axes, along with the best matching simulated patterns of the candidate structures and Mg. The latter is needed for the analysis as the experimental patterns contain a non-negligible contribution from the matrix. Comparison of the experimental and simulated patterns and considering the spot positions only illustrates that the two ternary precipitate patterns can be matched by either $Ca_2Mg_6Zn_3$, $Ca_2Mg_5Zn_5$, or $Ca_3Mg_{11}Zn_4$ for the zone axes [0001] (Figure 7, top row) and [$2\bar{1}\bar{1}0$] (Figure 7, bottom row), with a Mg matrix oriented along [$2\bar{1}\bar{1}0$] in both cases. However, when considering spot intensities, the [0001] pattern, in particular the stronger spots surrounding the transmitted beam making a hexagon, can best be fitted with the $Ca_2Mg_6Zn_3$ pattern. Besides, our experimental pattern perfectly matches the one in Fig. 5b of the paper by Jardim *et al.* [32]. The [$2\bar{1}\bar{1}0$] pattern shows also a relatively better fit to the $Ca_2Mg_6Zn_3$ pattern but the difference to the others is less obvious; a closer scrutiny reveals though that the overall intensity of the horizontal lines of spots best matches the $Ca_2Mg_6Zn_3$ pattern. It should be noted that the experimental pattern presents spots positions that are slightly shifted relative to the simulated patterns, which indicates that the structure may contain defects such as planar ones, possibly stacking faults. From this analysis it is deduced that the ternary precipitates in ZX20, heat-treated to 350°C, are $Ca_2Mg_6Zn_3$ and that the ternary crystal orientation relationships are the following: there is coherency with either $(0002)_{Ca_2Mg_6Zn_3} \parallel (2\bar{1}\bar{1}0)_{Mg}$ or $(0002)_{Ca_2Mg_6Zn_3} \parallel (0002)_{Mg}$ and $(2\bar{1}\bar{1}0)_{Ca_2Mg_6Zn_3} \parallel (2\bar{1}\bar{1}0)_{Mg}$.

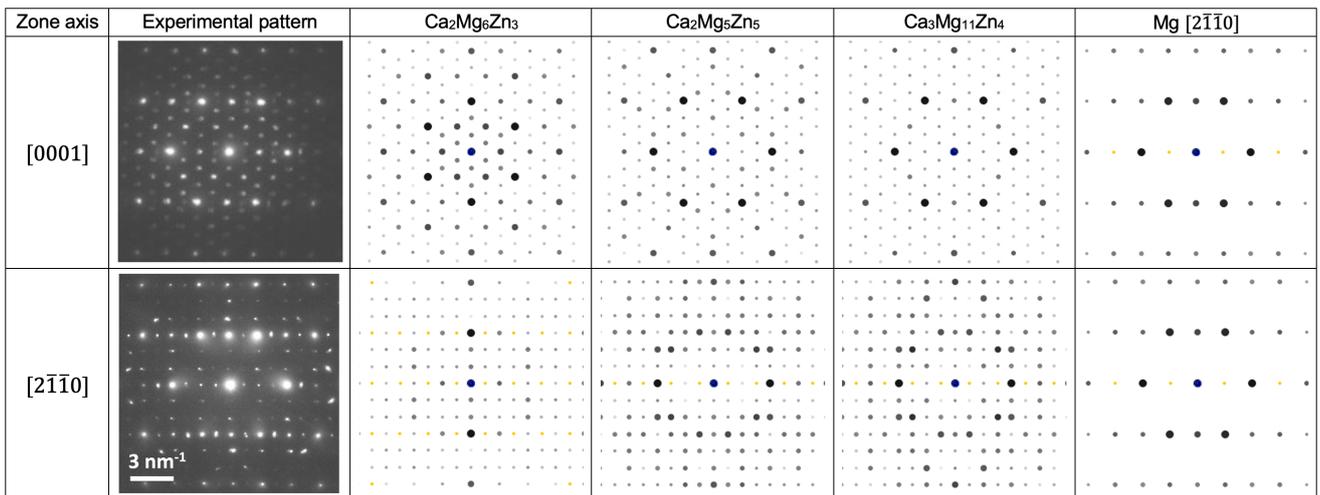

Figure 7. ZX20 heat-treated to 350 °C. Typical experimental nanodiffraction patterns obtained in the nanometric ternary precipitates in [0001] and [$2\bar{1}\bar{1}0$] directions, and corresponding best matching simulated patterns for $Ca_2Mg_6Zn_3$, $Ca_2Mg_5Zn_5$, $Ca_3Mg_{11}Zn_4$ and Mg structures. Note that the Mg matrix surrounding the precipitates contributes to the experimental pattern.

It thus appears that ZX20 steadily heated from a solid-solution condition up to 350 °C presents a majority of $Mg_2Ca$ precipitates and some ternary precipitates, which may be made of trigonal $Ca_2Mg_6Zn_3$. However, one may question whether the obtained structure is the equilibrium one because the heating rate of 20 K min$^{-1}$ is rather high and there was no hold time at 350 °C. The question arises then about the precipitation structure in a ZX20 alloy that is produced by hot extrusion at a constant temperature. Figure 8 presents combined Zn and Ca chemical maps of ZX20 heated to 350 °C (Fig. 8a), and ZX20 extruded at 330 °C shown at the same scale (Fig. 8b) and at a wider scale (Fig. 8c). Note that for the comparison to the static heat treatment to 350 °C an extrusion temperature of 330 °C was chosen as to account for the temperature increase due to deformation and friction during the extrusion process. It appears that following extrusion at 330 °C ZX20 presents also binary and ternary precipitates, as for the heat-treated ZX20, but with an inverted trend, i.e. with a majority of ternary precipitates. Moreover, they are larger, with an average size of 130 ± 13 nm, roughly equiaxed and faceted, and present a lower number density, of 1.5 (± 0.2) ×10$^{18}$ m$^{-3}$. About 10% of the precipitates are binary, presumably made of $Mg_2Ca$. They are not faceted, contrary to the ternary ones.

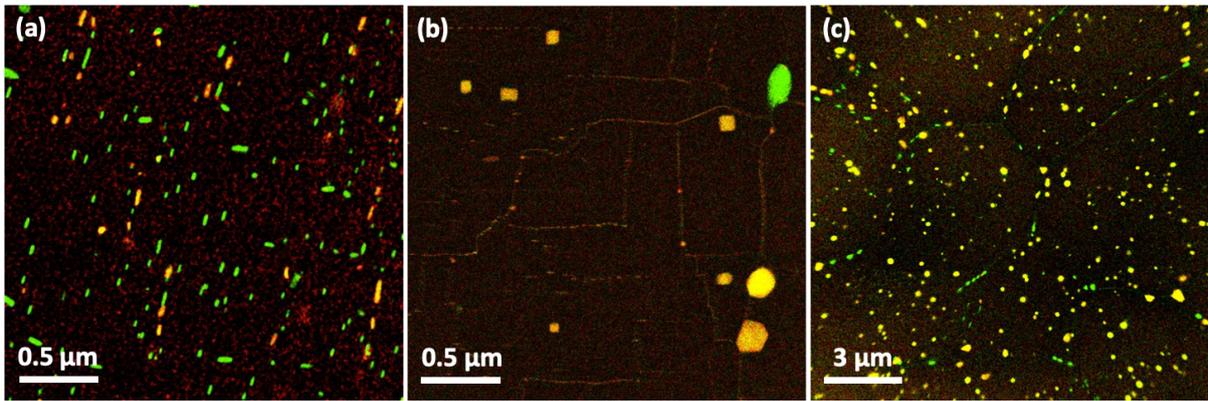

Figure 8. STEM EDS chemical maps of (a) ZX20 heat-treated to 350 °C and (b, c) ZX20 extruded at 330 °C, revealing the dispersion of binary (green) and ternary (yellow) precipitates. Color code is green for Ca and red for Zn (Zn+Ca makes yellow).

Figure 9 presents two experimental nanodiffraction patterns from two different ternary precipitates in as-extruded ZX20, along with the best matching simulated patterns of $Ca_2Mg_6Zn_3$, $Ca_2Mg_5Zn_5$ and $Ca_3Mg_{11}Zn_4$. When considering only the spots' positions, both experimental patterns can be matched with either of the simulated structures for the zone axes $[2\bar{1}\bar{1}0]$ and $[4\bar{5}10]$, respectively. When also considering intensities, however, the $[2\bar{1}\bar{1}0]$ pattern cannot be fitted as well with the $Ca_2Mg_6Zn_3$ pattern as it can be with $Ca_2Mg_5Zn_5$ or $Ca_3Mg_{11}Zn_4$. The $[4\bar{5}10]$ pattern in terms of intensities could be matched nearly equally well with either of those latter structures but a closer scrutiny of the outer-spots' intensities reveals that $Ca_2Mg_5Zn_5$ provides the best fit.

Figure 9. As-extruded ZX20. Typical experimental nanodiffraction patterns obtained in the nanometric Ca–Zn-rich precipitates in $[2\bar{1}\bar{1}0]$ and $[4\bar{5}10]$ directions, and the corresponding best matching simulated patterns for $Ca_2Mg_6Zn_3$, $Ca_2Mg_5Zn_5$ and $Ca_3Mg_{11}Zn_4$ structures.

Atomically resolved STEM imaging using the HAADF mode (Fig. 10) shows that the Zn–Ca rich precipitates (Fig. 10a) present facets on the prismatic planes (Fig. 10b). The partial power spectrum of Fig. 10b realized in the Mg matrix (Fig. 10c) and the one in the precipitate (Fig. 10d) reveal that the precipitates do not have a specific coherency relationship with the Mg matrix.

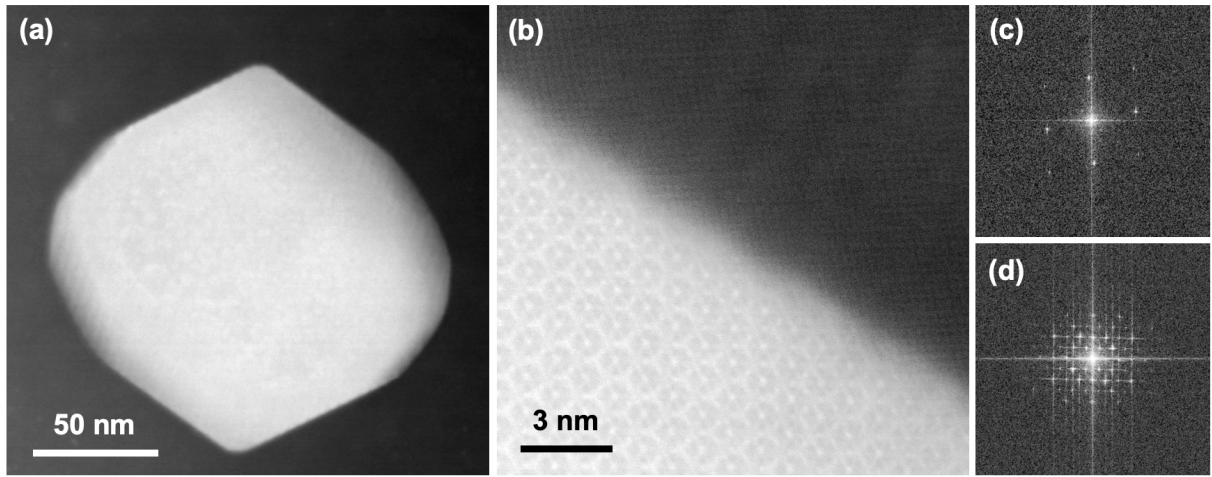

Figure 10. (a) HAADF STEM micrograph of a typical ternary precipitate found in ZX20 extruded at 330 °C. (b) Its atomic lattice revealed at higher magnification. (c, d) Image power spectrum revealing (c) in the matrix a $[10\bar{1}1]$ orientation and (d) in the precipitate a $[0001]$ orientation.

Ultrahigh-resolution HAADF STEM imaging and EDS chemical mapping were performed to clarify the structure of the ternary precipitates in as-extruded ZX20. Figure 11 shows two typical instances obtained in this way, taken along $[0001]$ (Fig. 11a-d) and $[2\bar{1}\bar{1}0]$ (Fig. 11e-h). Figure 11a shows a STEM contrast making a periodic hexagonal pattern. STEM-EDS (Fig. 11d) reveals that the hexagons' rim and center are principally made of Zn. Chemical maps further show that the inside of the hexagons, excluding the center, is rich in Ca (Fig. 11b), while Mg is distributed homogeneously but not present in the hexagons' center (Fig. 11c). A visual comparison of the candidate structures viewed along $[0001]$ (see Fig. S1) indicates that $Ca_2Mg_5Zn_5$ and $Ca_3Mg_{11}Zn_4$ may provide a good match, because the hexagonal pattern is met with both unlike with $Ca_2Mg_6Zn_3$, which is thus excluded. Moreover, the Zn map with the hexagonal pattern is visually better matched with the $Ca_2Mg_5Zn_5$ structure. Its projection overlaid on the STEM image in Fig. 11a shows indeed a good match in the contrast pattern and dimensions. Along $[2\bar{1}\bar{1}0]$, the structure in the HAADF STEM image reveals wavy spot lines (Fig. 11e), which appear in the chemical map to be rich in Zn (Fig. 11h). Some of the Zn-rich spots coincide with Ca-rich ones (Fig. 11f), while Mg is found to be nearly evenly distributed but not present in Ca-rich spots. The potentially best matching candidate structure is again $Ca_2Mg_5Zn_5$, as evaluated with the Zn map, but the signal is weak. Its projection along $[2\bar{1}\bar{1}0]$ overlaid on the STEM image is shown in Fig. 11e.

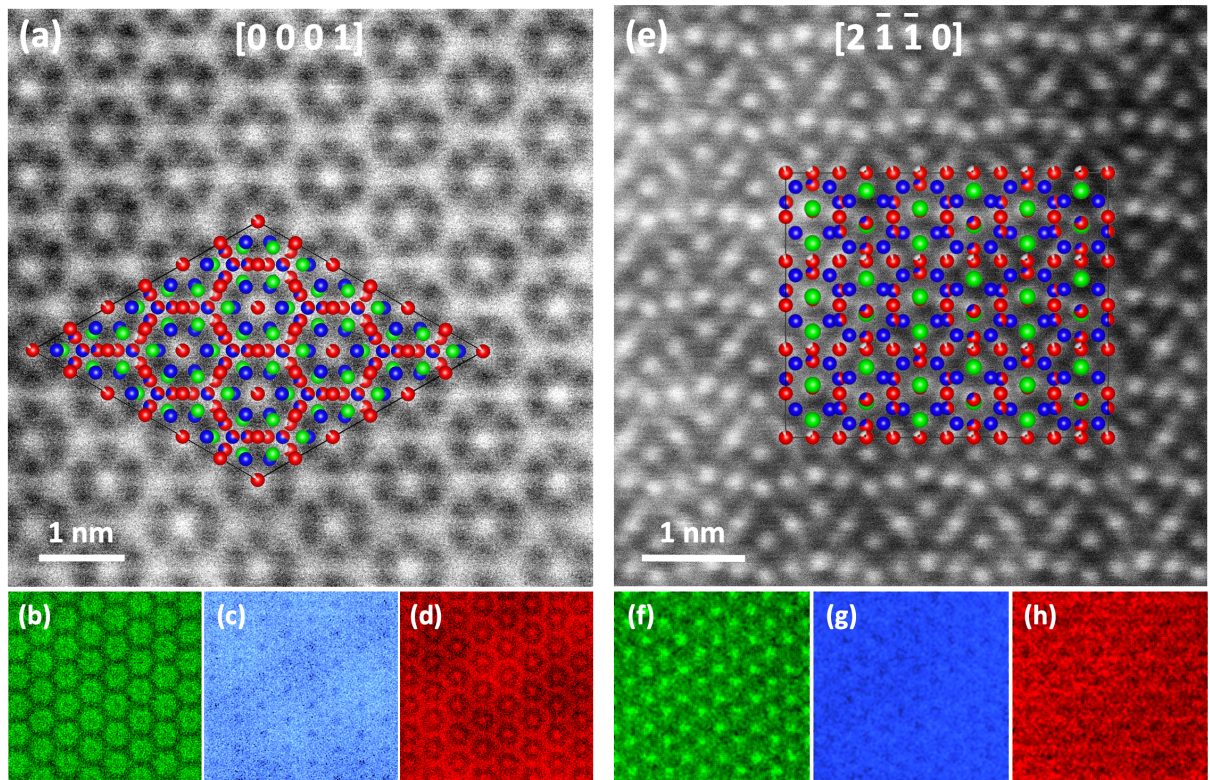

Figure 11. (a, e) Typical experimental ultrahigh-resolution HAADF STEM images and (b-d, f-h) corresponding EDS chemical maps of ternary precipitates in ZX20 extruded at 330 °C. The superimposed lattice structure in the STEM images is the one of $Ca_2Mg_5Zn_5$, the one best matching both the STEM images and the chemical maps (see details in the text). The crystal orientation is (a-d) [0001] and (e-h) $[2\bar{1}\bar{1}0]$. Color code: green for Ca, blue for Mg, and red for Zn.

While there are strong indications from the HAADF STEM imaging that $Ca_2Mg_5Zn_5$ is the structure of the Ca–Mg–Zn ternary precipitates present in ZX20 extruded at 330 °C, there remains doubt because of the poor fit to the chemical maps. We have thus performed a detail analysis, as presented in the Supplementary Information. First, simple 2D projections of the atoms of $Ca_2Mg_5Zn_5$ and $Ca_3Mg_{11}Zn_4$ structures were made to generate images that represent the experimental HAADF STEM and chemical maps along [0001] and $[2\bar{1}\bar{1}0]$ (Fig. S2 and S3, Supplementary Information). They indicate that $Ca_2Mg_5Zn_5$ is the structure that provides the best match, but chemical maps again generated a doubt on this conclusion. Considering that a simple 2D projection does not account for the real electron-matter interaction, we have then conducted simulations of the HAADF STEM images, using the THEMIS microscope's experimental parameters. The first step was to exclude image variations induced by the software choice. By comparing three of them (Fig. S4, Supplementary Information), it appeared that all yielded the same contrast type. The second step was to investigate the contrast variations we observed across all experimental HAADF STEM images, as they may stem from a structure variation within the same precipitate (Fig. S5, Supplementary Information), as indicated by the compound $Ca_3Mg_xZn_{15-x}$ ($4.6 \leq x \leq 12$) that has the same symmetry $P6_3/mmc$ [36] and whose composition range encompasses $Ca_2Mg_5Zn_5$ and $Ca_3Mg_{11}Zn_4$. The simulations clearly indicate that those image variations relate to a change in thickness, and not to a structure change. In sum, it appears that HAADF STEM images with the support of simulations can be reliably used to investigate the ternary structure. In addition, the precipitate structure seems to be constant among each other.

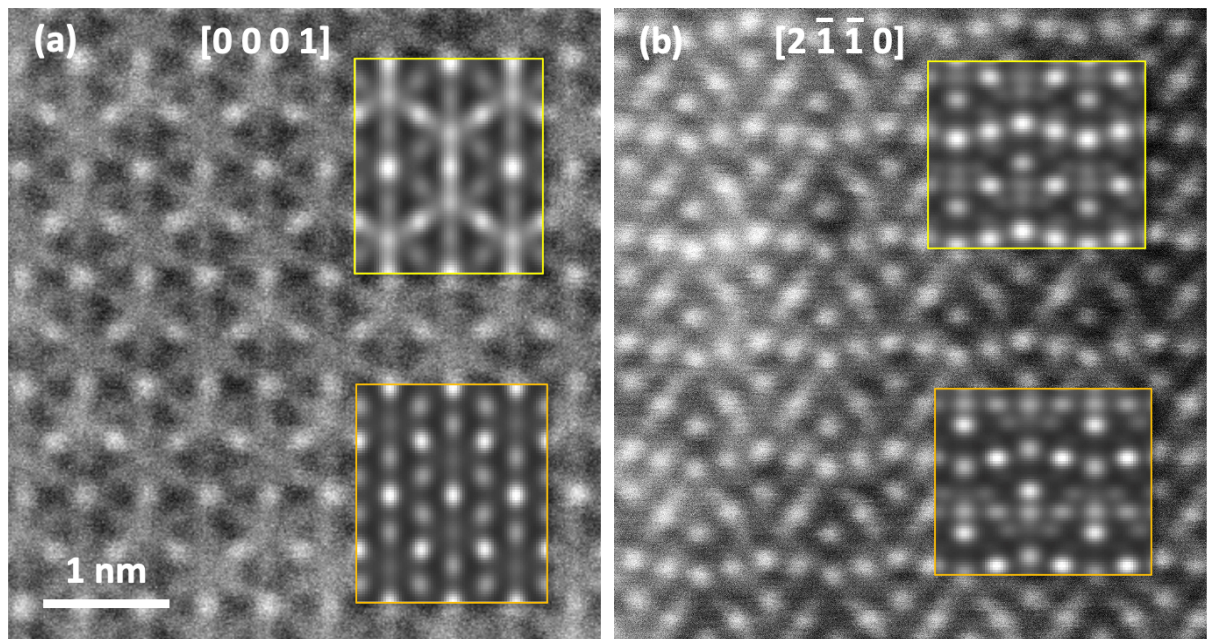

Figure 12. Typical ultrahigh-resolution HAADF STEM micrographs taken along (a) [0001] and (b) [2$\bar{1}\bar{1}$0] in Ca–Mg–Zn ternary precipitates present in ZX20 extruded at 330 °C. The inserts in the images show the best matching simulated images using either (top insert) the $Ca_2Mg_5Zn_5$ or (bottom insert) the $Ca_3Mg_{11}Zn_4$ structure.

Fig. 12 shows typical ultrahigh-resolution HAADF STEM micrographs taken along [0001] (Fig. 12a) and [2$\bar{1}\bar{1}$0] (Fig. 12b) in Ca–Mg–Zn ternary precipitates present in ZX20 extruded at 330 °C, with the inserts revealing the corresponding images simulated with either the $Ca_2Mg_5Zn_5$ or $Ca_3Mg_{11}Zn_4$ structure. From the visual inspection of the images we can conclude that, unlike $Ca_3Mg_{11}Zn_4$, $Ca_2Mg_5Zn_5$ allows for an excellent qualitative match to the experimental images. Thus, we infer that the ternary precipitates in as-extruded ZX20 are unambiguously of type $Ca_2Mg_5Zn_5$. On the other hand, the often-reported structure of $Ca_2Mg_6Zn_3$ (Fig. S1, Supplementary Information), which was identified here as the best matching candidate for the ternary precipitates in the material heat-treated with a rate of 20 K min$^{-1}$ to 350 °C (Fig. 7), can be clearly excluded for the material extruded at 330 °C.

# 4. Discussion

In the following we discuss the precipitation sequence in ZX20, which can be derived from the observations made by TEM and APT in the ZX20 alloy that was heated at 20 K min$^{-1}$. The phase composing the ternary precipitates observed at 350 °C is compared to the one observed in ZX20 extruded at 330 °C. It is then discussed in view of the phases predicted by thermodynamic calculations. Finally, a relationship between precipitation and hardness is established.

## 4.1 Precipitation structure

In the specimen heat-treated to 205 °C the analysis of the APT measurements revealed clustering of Ca, Zn and Zn–Ca (Fig. 3b), which is not visible in TEM. Ca is nearly insoluble in Mg up to about 200 °C [54], while Zn has some solubility (~1 at.% at 200 °C [55]). Moreover, in Mg, Ca diffuses about 3 times faster than Zn [56]. These characteristics indicate that Ca can form clusters first from a kinetic point of view. This assumption is supported by the fact that the observed smaller clusters (~2.6 nm), which are assumed to be the primary and less mature clusters, are in majority Ca-rich (about 30% of the clusters observed) and have a lower Zn content relative to the matrix (Fig. 3c, top). The larger clusters (~3 nm), indicative of maturation, make the majority (at ~48%) and are Zn–Ca-rich (Fig. 3c, bottom) with a Zn-to-Ca ratio of about 2 (Table 2). The co-presence of Ca and Zn can be explained by (1) the difference in the atomic radii of Zn and Ca relative to Mg and (2) their difference in mixing enthalpies. Indeed, incorporating the undersized Zn atoms to the clusters of oversized Ca atoms (relative to the Mg atoms) can favorably reduce the lattice mismatch [57, 58] and thus facilitate the growth of precursor Ca-rich clusters. In addition, with a mixing enthalpy of Ca–Zn amounting to -22 kJ mol$^{-1}$, which is one order of magnitude larger than the (also negative) mixing enthalpy of Mg–Zn and Mg–Ca [3], there is a strong attractive interaction between Ca and Zn. Zn would thus tend to segregate to the primary Ca clusters. Note that the measured content of Zn and Ca in the matrix is respectively 0.47 at.% and 0.14 at.%, while the nominal content in the matrix is respectively 0.56 at.% and 0.15 at.%. There is thus a slight decrease in alloying element content in the matrix due to precipitation, but the Ca content in solution remains high relative to its stated solubility limit at this temperature.

After heat treatment to 260 °C, a high number density of nanometric precipitates is observed on the basal plane, which are GP zones as revealed by HR-TEM (Fig. 4). APT and EDS in STEM reveal that they are enriched in Zn and Ca; in the APT 3D-reconstruction they exhibit an elliptical disk shape while in TEM and in STEM they appear as sharp thin disks, one or a few atomic planes thick. They contain Mg, Zn and Ca with a Zn-to-Ca ratio of about 1.2 as obtained from the APT measurements (Table 2). EDS indicates a much larger ratio (3.3±3.0, see Table 2), noting that with EDS it is only possible to obtain the ratio of Zn and Ca but not that to Mg, because the surrounding Mg matrix biases the measured Mg content in the precipitate. In addition, the measured Zn content is biased by Zn redeposition on the free surfaces of the TEM thin foil during ion beam milling, a redeposition that is more intense than for Ca and Mg due to its higher content in the alloy (compared to Ca) and its higher atomic weight (compared to Ca and Mg). This leads to an overestimation of the Zn content that becomes larger the thinner the area, which unfortunately is where GP zones can be best observed. EDS quantification is thus not considered for the GP zones. Nevertheless, the comparison of the Zn-to-Ca ratio in the GP zones of about 1.2, as obtained by APT, to the nominal one of 3.7 for the ZX20 alloy indicates again that Ca clusters more strongly than Zn, which supports our hypothesis of preferential clustering of Ca because of its low solubility and faster diffusivity.

It should be noted that composition measurements by APT may be biased by known artefacts: on the one hand, there is the difference in the evaporation field between Zn, Ca and Mg, where Zn is the hardest to evaporate (33, 18, and 21 V nm$^{-1}$, respectively). Preferential retention of Zn may therefore lead to a Zn amount measured in a cluster that is first locally slightly lower than the real amount and then higher once the cluster has been passed. This relates to the slight shift observed between the isoconcentration surfaces for Zn and Ca around the position of the Zn–Ca one, and the fact that the orientation of the observed Zn-rich and Ca-rich side of the Zn–Ca cluster is the same for all precipitates (Fig. 4e). This orientation would thus relate to the evaporation direction, with a Zn distribution seemingly lagging behind the Ca one. On the other hand, there is a local magnification effect [59], which artificially decreases the local concentration of the detected elements with a spread of their spatial distribution. This artefact is a common challenge in APT, already observed for GP zones in a lean Mg–Al–Ca–Mn alloy [60], and needs to be taken into consideration. Due to the thin-plate shape of the precipitates and the local magnification effects, Mg atoms from the matrix are included in the reconstruction of the precipitates, leading to a large overestimation of the Mg concentration. Because this impedes obtaining an absolute measurement of the Zn and Ca concentration composing the GP zones, only their Zn-to-Ca ratio has been considered.

While STEM imaging (Fig. 4a) indicates that the precipitates could be few atomic layers thick, HRTEM imaging (Fig. 4d) shows that most of them are actually monolayered. The former is consistent with the observation of Oh-ishi *et al.* [3], who reported for an Mg-1.6Zn-0.3Ca alloy GP zones that are at most multiple atomic planes thick. These thin precipitates have been identified as $Ca_2Mg_6Zn_3$ by electron diffraction, but without chemical information [3]. Assuming that the GP zones observed in the present study have the $Ca_2Mg_6Zn_3$ structure, the Zn-to-Ca ratio would be 1.5. Thus, while the absolute concentration in Ca, Mg and Zn measured by APT does not fit the $Ca_2Mg_6Zn_3$ structure, the measured Zn-to-Ca ratio of 1.2 ± 0.8 presents a good match. The large discrepancy of the absolute concentrations can be explained by the above-mentioned local magnification effects in APT, which leads to a large overestimation of the Mg content. The GP zones in the present study might thus come close to the structure of $Ca_2Mg_6Zn_3$, although the GP zones in Oh-ishi's work may present a more mature structure, closer to equilibrium than the ones observed here in ZX20, because they have more than one atomic layer. Note that the matrix contains 0.36 at.% Zn and 0.11 at.% Ca, which compared to the 205 °C heat-treated case marks a further decrease of the alloying element content in the matrix due to precipitation. The Ca content in solution remains high but this is expected as the temperature is higher.

After the heat treatment to 350 °C two distinct precipitate types were observed, namely small and large binary ones and small and large ternary ones (Fig. 5). Most binary and ternary precipitates are coherent with the matrix, having in common the basal plane of the Mg structure. TEM EDS was deemed reliable compared to the case of the GP zone analysis because the areas of interest are much thicker and thus quantification is less affected by Zn redeposition.

The large, thus more mature, binary precipitates are assumed to be made of $Mg_2Ca$, following Ref. [2]. The Zn-to-Ca ratio of 0.20 measured in these precipitates by EDS varies largely with a maximum at 0.34 (Table 2), which corresponds well with the maximum solubility of Zn in $Mg_2Ca$ found for the compound $Ca_{33.3}Mg_{55.9}Zn_{10.8}$ [38, 39], giving a ratio of 0.324. The large ternary precipitates have a Zn-to-Ca ratio of about 1.5 (Table 2). This together with the nanodiffraction pattern analysis (Fig. 7) indicates that they are mature precipitates made of the $Ca_2Mg_6Zn_3$ phase suggested by Jardim *et al.* [32]. The small, binary precipitates contain quite some Zn, with a Zn-to-Ca ratio of 0.6, which is twice the reported maximum Zn solubility in $Mg_2Ca$ of 10.8 at.% that gives a ratio of 0.324 (see above). This indicates that the assumption of a $Mg_2Ca$ structure for the small binary precipitates may be too crude. Conversely, the small ternary precipitates present a Zn-to-Ca ratio of 1.4, which is close to the one of the $Ca_2Mg_6Zn_3$ phase. It thus seems that the GP zones observed at 260 °C with a Zn-to-Ca ratio of 1.2 grow with a forking of their structure as temperature is increased. Starting with a structure close to the one of the ternary $Ca_2Mg_6Zn_3$ phase they either lose Zn and at the same time probably incorporate some Ca remaining in the matrix to form the binary $Mg_2Ca$ phase, or reinforce the ternary $Ca_2Mg_6Zn_3$ phase by incorporating Zn remaining in the matrix.

It is interesting to note that in the sample heat-treated at 260 °C the precipitates appear rather homogeneously distributed in size and space, with GP zones being a few nanometers in size and having a Zn-to-Ca ratio (Table 2) of about 1.2. In contrast, in the sample heat-treated at 350 °C the precipitates' population reduced due to segregation into either binary or ternary precipitates that are tens of nanometers in size but still coherent with the matrix. It also appears that the binary precipitates induced compression in the surrounding lattice (Fig. 6c). For some of the larger binary ones punching of dislocation loops also occurred (Fig. 5a), indicative of a massive matrix deformation. Conversely, the ternary precipitates seem to blend in the Mg matrix without considerable deformation of the host lattice (Fig. 6b) or, if any, it shows up as an expansion of the matrix in the precipitate (Fig. 6d). When considering that either of these phases can be constructed from the Mg hexagonal basal plane via substituting Mg atoms by Zn or Ca and, further, that Ca atoms are larger than Mg ones while Zn atoms are smaller, the $Mg_2Ca$ phase will occupy a larger volume than the Mg matrix [2]. This leads to compression of the surrounding lattice, while, in contrast, the $Ca_2Mg_6Zn_3$ with a majority of Zn over Ca will occupy a volume that is smaller [31], promoting an expansion of the surrounding lattice.

## 4.2 Thermodynamic calculation and pathways of precipitation kinetics

Let us compare the microstructural information obtained in this study with the thermodynamic calculation presented in Fig. 1. In the material heat-treated at 205 °C, while the precipitation of Zn-rich, Ca-rich and Zn–Ca-rich nanometric clusters is not mature enough to give clearly identified phases, one can nonetheless state that there is a tendency of ternary compound formation, as such clusters dominate in number density, consistent with Fig. 1. Note that in no case did we identify the IM3 phase seen in Fig. 1. In the material heat-treated at 260°C we observed precipitates (GP zones) that are made of a ternary phase, possibly a precursor to $Ca_2Mg_6Zn_3$, again consistent with Fig. 1, while at 350 °C precipitates are either made of the binary $Mg_2Ca$ phase or the ternary $Ca_2Mg_6Zn_3$ phase, with the former dominating, which does not correspond to the results of Fig. 1. However, the temperature dependence of the volume fraction shows a consistency with the overall trend indicated by the calculation: with increasing temperature the volume fraction of $Ca_2Mg_6Zn_3$ decreases and past a certain temperature (about 365 °C) $Mg_2Ca$ appears. Its volume fraction increases with increasing temperature, eventually exceeding the one of $Ca_2Mg_6Zn_3$, which disappears at about 380 °C. In the calculation $Mg_2Ca$ dominates from about 375 °C, a situation that corresponds to a temperature that is 25 °C lower than in the present experimental case. Note, however, that the calculations assume thermodynamic equilibrium, a state that cannot be expected in the heat-treated samples due to their short annealing time. On the other hand, one would expect that in order to reach equilibrium faster, the temperature should be increased, which goes against the observation: equilibrium seems to be reached at a lower temperature. This indicates that the calculation does not reproduce the transition temperatures between the phases accurately. Note further that the calculated solidus temperature is 460 °C while the measured one is 560 °C, which is an inconsistency of the thermodynamic calculation that was also revealed in the case of an Al alloy [61].

In the ZX20 alloy extruded at 330 °C the situation is different: before extrusion, the material was annealed at 250 °C for 30 minutes, which, compared to ZX20 heat-treated to 260 °C, brings the system closer to thermodynamic equilibrium. At 250 °C, according to thermodynamic calculations (Fig. 1), precipitates should be made exclusively of the ternary phase $Ca_2Mg_6Zn_3$. The extended time at 250 °C may have allowed the precipitates to reach such an equilibrium phase, in contrast to the heat-treated condition where the heating rate of 20 K min$^{-1}$ provided only a brief exposure to the studied temperatures. After extrusion to 330 °C, $Ca_2Mg_6Zn_3$ was, however, clearly excluded (Figs. 11 and 12), revealing instead the ternary phase $Ca_2Mg_5Zn_5$. Besides, the precipitates were incoherent and larger than in the material heat-treated to 350 °C. Here it should be noted again that although the material was extruded at 330 °C the effective temperature may have reached 350 °C because of the large plastic deformation and friction involved in the extrusion process, for a time that probably has extended to minutes. In this light it may be surprising that the majority of the precipitates observed are still made of the ternary phase, as, in view of the results of ZX20 heat-treated to 350 °C, one can expect binary $Mg_2Ca$ precipitates to dominate. However, the precipitates may have reached already during the 250 °C anneal a ternary equilibrium phase ($Ca_2Mg_5Zn_5$) that is more stable than $Ca_2Mg_6Zn_3$. This, coupled to the fact that they are incoherent and larger ($\approx$130 nm) than the $Ca_2Mg_6Zn_3$ precipitates of the 350 °C heat-treated material, make them more difficult to dissolve or to change their structure.

The dichotomy between the crystallographic structure of the binary $Mg_2Ca$ phase and the different ternary phases ($Ca_2Mg_6Zn_3$ and $Ca_2Mg_5Zn_5$) can be related to the spatial arrangement of the Ca atoms that start in the Mg lattice. In the present study it was observed that the coherent $Mg_2Ca$ and $Ca_2Mg_6Zn_3$ precipitates share with the Mg matrix the basal plane. In the basal plane of $Mg_2Ca$ the Ca atoms make hexagons, which relate to the ones found in the Mg structure and is also the case for the trigonal structure of the $Ca_2Mg_6Zn_3$ phase. Conversely, in the basal plane of the hexagonal structure of $Ca_2Mg_5Zn_5$ the Ca atoms make triangles and irregular hexagons that do not relate to the Mg-matrix basal atomic arrangement.

The precipitation upon continuous heating can thus be rationalized by the following sequence. Starting from Zn and Ca in solid solution with Mg, Ca is the first to precipitate upon heating to 205 °C, because of its lower solubility and higher diffusivity relative to Zn, forming nanometric clusters. Zn also tends to precipitate but to a lesser extent and mostly co-precipitates with Ca, as this relieves lattice mismatch. The formation enthalpy of Zn–Ca also favors their mixing, leading to ternary clusters. Upon further heating to 260 °C monolayered GP zones form that have a composition tending towards the one of $Ca_2Mg_6Zn_3$, whose structure can be built through the hexagonal structure of the Mg matrix via an arrangement of Zn and Ca atoms on the shared basal plane. With increasing temperature, they grow and thicken into a well-defined coherent $Ca_2Mg_6Zn_3$ structure. Some of them may dissolve, as indicated by the modest endothermic peak at about 270 °C (Fig. 2). According to the thermodynamic calculation, the volume fraction of $Ca_2Mg_6Zn_3$ decreases with increasing temperature when approaching the $Mg_2Ca$ field, indicating that $Ca_2Mg_6Zn_3$ becomes destabilized. Here, Zn may dissolve from the $Ca_2Mg_6Zn_3$ precipitates and the Ca remaining in solution may replace it to form $Mg_2Ca$. Again, this is consistent with a ternary phase made of the trigonal phase $Ca_2Mg_6Zn_3$: as $Ca_2Mg_6Zn_3$ and $Mg_2Ca$ are coherent with the Mg matrix the replacement of the Zn atoms by Mg or Ca to form $Mg_2Ca$ does not imply a change of the crystal structure as drastic as it would occur in precipitates made of incoherent $Ca_2Mg_5Zn_5$.

One can extrapolate this crystallographic reasoning to the case of as-extruded ZX20: the equilibrium ternary phase is $Ca_2Mg_5Zn_5$, which dominated over the binary phase $Mg_2Ca$ despite the fact that the calculated phase diagram indicates the opposite. This shows that this ternary phase does not transform easily to the binary phase, contrary to the $Ca_2Mg_6Zn_3$ phase. In addition to the arguments invoked earlier (incoherence, larger size and potentially higher phase stability), the difficulty to generate a phase transition may simply relate to the larger structural difference between the $Ca_2Mg_5Zn_5$ structure and coherency to the matrix compared to the one between coherent $Mg_2Ca$ and $Ca_2Mg_6Zn_3$.

### 4.3 Precipitation hardening

In the following the impact of precipitation on hardness response upon heating is discussed. Precipitation hardening occurs because dislocations, i.e. the vectors of plasticity, are impeded in their glide by the precipitates. Dislocations bow out between the precipitates and may eventually pass them by shearing. If the precipitates are unshearable, the dislocations may cross-slip to go over or below the precipitates or circle around them, forming an Orowan dislocation loop. Cross-slipping is hardly possible in Mg's hexagonal structure as the shear stress for dislocations to cross-slip is 100 times the stress needed to glide on the basal plane. For shearable obstacles in high number density the following equation [62] can be used to express hardening or an increase in tensile yield strength, $\Delta\sigma$:

$$\Delta\sigma = \sqrt{\frac{3}{4\pi\beta}} \frac{k^{\frac{3}{2}} M\mu}{\sqrt{b}} \sqrt{f\frac{d}{2}}, \tag{1}$$

where $\beta$ is a parameter equal to 0.5 [62], $k$ is a measure of the precipitate strength to hinder dislocation passage, $M$ is the Taylor factor accounting for polycrystallinity, $\mu$ is the matrix shear modulus, $b$ is the magnitude of the Burgers vector, $f$ is the volume fraction of the precipitates derived from their number density $N$ and their volume, and $d$ is the dimension of the precipitate in the glide plane. The hardening due to unshearable obstacles is given by [63]:

$$\Delta\sigma = M\alpha\mu b\sqrt{Nd}, \tag{2}$$

where $\alpha$ is the obstacle strength, reaching 1 for unshearable obstacles. For Mg, $\mu$ is 17 GPa, $b$ is 0.32 nm, and $M = 3$. However, in our study we consider that the hardness was assessed in single-crystalline regions because the grain size was larger than the indenter imprint, so that $M = 1$ is considered. To convert yield stress to hardness, the measured hardness is multiplied by 3, as a rule of thumb, and then divided by 9.807 for the unit change [63].

After heating to 205 °C the hardness increases by about 4.6 HV or about 10% (Fig. 2), showing that the clustering observed at 205 °C induces some hardening. Extended clustering was observed, amounting to $5.6 \times 10^{23}$ m$^{-3}$ in total, considering all Zn-rich, Ca-rich and Zn–Ca-rich clusters (Fig. 3a). The size of the clusters is on average 2.9 nm and their volume fraction is 0.0073 when spherical geometry is assumed. The lack of any visible strain around them, as observed by TEM, and their cloud-like appearance in APT indicates that their obstacle strength is weak and that they are thus shearable. With this in mind Eq. 1 was considered, which delivered an obstacle strength $k$ of 0.0367 to match the measured hardening.

When passing 260 °C, hardening becomes significant with an increase of about 9 HV (= 23%) compared to the solution-heat-treated condition. This peak hardness corresponds to the end of the second DSC peak (Fig. 2). Microstructural analysis revealed that the hardening relates to the formation of GP zones, which are about 8 nm in size and appear in a density of $1.3 \times 10^{23}$ m$^{-3}$, which is about 4 times lower than the one of the clusters observed at 205 °C. These GP zones are atomically thin and lie on the glide plane of dislocations. They cannot be *stricto sensu* sheared but instead the shearing can occur in their surrounding strain field as seen in TEM (Fig. 4d). It is the difference in atomic nature, together with the surrounding strain field, that makes GP zones obstacles to dislocation glide (pp. 376-379 in Ref. [64]). The volume of the GP zones is considered as a disk of 7.8 nm in diameter that has a thickness corresponding to the Mg-interplanar distance along the $c$ axis of 0.26 nm [60], yielding a volume fraction of 0.0016. The analysis of the hardening using Eq. (1) results in an obstacle strength $k$ of 0.0685 to match the measured hardness, which shows that GP zones are stronger obstacles than the nanometric Zn-rich, Ca-rich and Zn–Ca-rich clusters, but still weaker than the GP zones in the AXM-lean alloy presented in Ref. [56], where a $k$ factor of 0.161 was determined.

Upon heating to 350 °C the hardness shows a strong decrease, with a net-hardening gain of 3 HV compared to the solid-solution starting condition. This temperature corresponds to the end of the third DSC peak and to the state where larger precipitates made of either Mg$_2$Ca or Ca$_2$Mg$_6$Zn$_3$ are observed. Their number density is in total $2.9 \times 10^{20}$ m$^{-3}$ (Table 2, considering the small and large precipitates) and thus about 2000 and 450 times lower than the one of the clusters observed at 205 °C and of the GP zones observed at 260 °C, respectively. The precipitates are cylinders with an average length of 56 nm, which is taken as their dimension $d$ because they are elongated in the glide plane (Fig. 6b). Owing to their low density, their larger size, the surrounding strain field around the $Mg_2Ca$ precipitates observed in TEM (e.g. Fig. 5a), and the reported brittleness of the $Ca_2Mg_6Zn_3$ phase [27], they are first considered unshearable. Eq. 2 was thus used to estimate their contribution to hardening and $\alpha$ was initially taken as 1, corresponding to unshearable obstacles, to account for the maximum possible hardening. With this assumption, Eq. 2 yields a hardening of 6.7 HV, which is about twice the measured one for the heat-treated condition at 350 °C. To match the measured hardening the obstacle strength $\alpha$ was then adjusted to 0.45, which still makes the precipitates strong obstacles but somewhat shearable, as one can envisage from their coherency with the matrix.

The main reason for the over-aging is thus the strong decrease in the number density of precipitates relative to the GP zones. This low hardening correlates well with previous studies, which showed that in particular the $Mg_2Ca$ phase, dominating $Ca_2Mg_6Zn_3$ in number density and in strength of the surrounding strain field, has a moderate effect on hardening [2].

## 5. Summary and Conclusions

Precipitation in Mg-1.5Zn-0.25Ca (ZX20, in wt.%) has been investigated by DSC, APT and TEM, equipped with EDS for chemical mapping down to the atomic scale, following a linear heating of the material at a rate of 20 K min$^{-1}$ from the solid-solution state at RT up to 500 °C. Vickers hardness was measured to assess the impact of precipitation on hardness. The structures of the precipitates generated by the linear heating were compared to the ones observed in ZX20 extruded at 330 °C. The following conclusions are drawn:

- DSC performed at a rate of 20 K·min$^{-1}$ from RT to 500 °C on solution-heat-treated ZX20 alloys presented exothermic peaks at 125, 250 and 320 °C, which ended at about 205, 260 and 350 °C, respectively.
- After heating to 205 °C, APT revealed clustering in the form of Ca-rich, Zn-rich or Zn–Ca-rich globular clusters of about 3 nm or smaller in size and with a density of $5.7 \times 10^{23}$ m$^{-3}$. The Zn–Ca-rich clusters made about half of the cluster population, were the largest, and had a Zn-to-Ca ratio of about 2. Ca-rich clusters dominated the smaller ones with a population of 30% and a size of about 2.6 nm.

- After heating to 260 °C, APT and TEM revealed the presence of atomically thin Guinier–Preston zones of about 8 nm in size and with a number density of $1.3 \times 10^{23}$ m$^{-3}$, containing a ternary phase with a Zn-to-Ca ratio of 1.2. It is related to the trigonal ternary $Ca_2Mg_6Zn_3$ phase.

- After heating to 350 °C, TEM revealed larger, elongated precipitates of up to about 100 nm in length with a number density of $2.9 \times 10^{20}$ m$^{-3}$. They are coherent with the matrix and made of two different phase types: either a binary Ca-rich one assumed to be made of the $Mg_2Ca$ phase, which dominates in number density, or a ternary Zn–Ca-rich one.

- The debated crystallographic structure of the ternary phase has been analyzed by TEM nanodiffraction: it is best matched by $Ca_2Mg_6Zn_3$, which provides also the best match for the Zn-to-Ca ratio of about 1.5 obtained by EDS.

- In ZX20 extruded at 330 °C the precipitation microstructure is different from the one observed following linear heating. It is characterized by a majority of Zn−Ca-rich ternary precipitates and some Ca-rich ones, both being at most semi-coherent with the matrix. Following TEM-based nanodiffraction, atomically resolved HAADF STEM imaging, and chemical mapping, the ternary precipitates appear to be made of hexagonal $Ca_2Mg_5Zn_5$.

- This hexagonal $Ca_2Mg_5Zn_5$ structure is confirmed by HAADF STEM image simulations, using three different simulation software.

- As the hold times at the temperatures of interest in hot extrusion compared to the linear heating are longer, $Ca_2Mg_5Zn_5$ appears as an equilibrium phase that is more stable than $Ca_2Mg_6Zn_3$.

- Precipitation can be used in ZX20 for hardening. The hardness exhibits an increase with increasing temperature up to 290 °C, followed by a decrease beyond that temperature. More precisely, the hardness HV amounts to 39.3 in the solutionized condition, and to 43.9, 48.3 and 42.3 following heating to 205 °C, 260 °C and 350 °C, respectively.

- The largest contribution to hardening is provided by the fine and dense dispersion of Zn–Ca-rich GP zones. In contrast, the nanometric globular Zn-, Ca- and Zn–Ca-rich clusters provide only small hardening, although their number density can be high. Likewise, the larger binary and ternary precipitates occurring in lower number density generate only limited hardening.

# Acknowledgements

We thank T. Akhmetshina of LMPT, ETH Zurich, for discussions on the crystallography of the ternary phases. We are also grateful to the ETH Zurich Scientific Center for Optical and Electron Microscopy (ScopeM) and the EPF Lausanne Interdisciplinary Center for Electron Microscopy (CIME) for providing access to their instruments. Financial support from the Swiss National Science Foundation via an SNF Sinergia grant (CRSII5-180367) is also gratefully acknowledged.

# References

[1] A.A. Luo, Recent magnesium alloy development for elevated temperature applications, Int Mater Rev 49 (1) (2004) 13-30.
[2] J.F. Nie, B.C. Muddle, Precipitation hardening of Mg-Ca(-Zn) alloys, Scripta Mater 37 (10) (1997) 1475-1481.
[3] K. Oh-ishi, R. Watanabe, C.L. Mendis, K. Hono, Age-hardening response of Mg-0.3 at.%Ca alloys with different Zn contents, Mat Sci Eng A-Struct 526 (1-2) (2009) 177-184.
[4] G.V. Raynor, The Physical Metallurgy of Magnesium and its Alloys, Pergamon Press, London, U.K., 1959.
[5] B.S. You, W.W. Park, I.S. Chung, The effect of calcium additions on the oxidation behavior in magnesium alloys, Scripta Mater 42(11) (2000) 1089-1094.
[6] O. Beffort, C. Hausmann, Magnesium Alloys and their Applications, Werkstoff-Informationsgesellschaft GmbH, Frankfurt, 2000.
[7] C.J. Bettles, M.A. Gibson, K. Venkatesan, Enhanced age-hardening behaviour in Mg-4 wt.% Zn micro-alloyed with Ca, Scripta Mater 51 (3) (2004) 193-197.
[8] T. Bhattacharjee, C.L. Mendis, K. Oh-ishi, T. Ohkubo, K. Hono, The effect of Ag and Ca additions on the age hardening response of Mg-Zn alloys, Mat Sci Eng A-Struct 575 (2013) 231-240.
[9] J.D. Robson, D.T. Henry, B. Davis, Particle effects on recrystallization in magnesium-manganese alloys: Particle pinning, Mat Sci Eng A-Struct 528 (12) (2011) 4239-4247.
[10] J. Hofstetter, S. Ruedi, I. Baumgartner, H. Kilian, B. Mingler, E. Povoden-Karadeniz, S. Pogatscher, P.J. Uggowitzer, J.F. Loffler, Processing and microstructure-property relations of high-strength low-alloy (HSLA) Mg-Zn-Ca alloys, Acta Mater 98 (2015) 423-432.
[11] M. Stefanidou, C. Maravelias, A. Dona, C. Spiliopoulou, Zinc: a multipurpose trace element, Arch Toxicol 80(1) (2006) 1-9.
[12] B. Zberg, P.J. Uggowitzer, J.F. Loeffler, MgZnCa glasses without clinically observable hydrogen evolution for biodegradable implants, Nat Mater 8 (11) (2009) 887-891.
[13] T. Kraus, S.F. Fischerauer, A.C. Haenzi, P.J. Uggowitzer, J.F. Loeffler, A.M. Weinberg, Magnesium alloys for temporary implants in osteosynthesis: In vivo studies of their degradation and interaction with bone, Acta Biomater 8 (3) (2012) 1230-1238.
[14] J. Hofstetter, M. Becker, E. Martinelli, A.M. Weinberg, B. Mingler, H. Kilian, S. Pogatscher, P.J. Uggowitzer, J.F. Loffler, High-Strength Low-Alloy (HSLA) Mg-Zn-Ca Alloys with Excellent Biodegradation Performance, JOM 66 (4) (2014) 566-572.
[15] B. Zhang, Y. Hou, X. Wang, Y. Wang, L. Geng, Mechanical properties, degradation performance and cytotoxicity of Mg-Zn-Ca biomedical alloys with different compositions, Mat Sci Eng C-Mater 31 (8) (2011) 1667-1673.
[16] Y. Sun, M.-x. Kong, X.-h. Jiao, In-vitro evaluation of Mg-4.0Zn-0.2Ca alloy for biomedical application, T Nonferr Metal Soc 21 (2011) S252-S257.

[17] S.Y. Cho, S.-W. Chae, K.W. Choi, H.K. Seok, Y.C. Kim, J.Y. Jung, S.J. Yang, G.J. Kwon, J.T. Kim, M. Assad, Biocompatibility and strength retention of biodegradable Mg-Ca-Zn alloy bone implants, J Biomed Mater Res B 101B (2) (2013) 201-212.
[18] P.-R. Cha, H.-S. Han, G.-F. Yang, Y.-C. Kim, K.-H. Hong, S.-C. Lee, J.-Y. Jung, J.-P. Ahn, Y.-Y. Kim, S.-Y. Cho, Biodegradability engineering of biodegradable Mg alloys: Tailoring the electrochemical properties and microstructure of constituent phases, Sci Rep 3 (2013) 2367.
[19] K. Pichler, S. Fischerauer, P. Ferlic, E. Martinelli, H.-P. Brezinsek, P.J. Uggowitzer, J.F. Loeffler, A.-M. Weinberg, Immunological Response to Biodegradable Magnesium Implants, JOM 66 (4) (2014) 573-579.
[20] F. Witte, V. Kaese, H. Haferkamp, E. Switzer, A. Meyer-Lindenberg, C.J. Wirth, H. Windhagen, In vivo corrosion of four magnesium alloys and the associated bone response, Biomaterials 26 (17) (2005) 3557-3563.
[21] J. Nagels, M. Stokdijk, P.M. Rozing, Stress shielding and bone resorption in shoulder arthroplasty, J Shoulder Elb Surg 12(1) (2003) 35-39.
[22] H.R. Bakhsheshi-Rad, E. Hamzah, A. Fereidouni-Lotfabadi, M. Daroonparvar, M.A.M. Yajid, M. Mezbahul-Islam, M. Kasiri-Asgarani, M. Medraj, Microstructure and bio-corrosion behavior of Mg-Zn and Mg-Zn-Ca alloys for biomedical applications, Mater Corros 65(12) (2014) 1178-1187.
[23] D. Zander, N.A. Zumdick, Influence of Ca and Zn on the microstructure and corrosion of biodegradable Mg-Ca-Zn alloys, Corros Sci 93 (2015) 222-233.
[24] M. Cihova, P. Schmutz, R. Schäublin, J.F. Loffler, Biocorrosion Zoomed In: Evidence for Dealloying of Nanometric Intermetallic Particles in Magnesium Alloys, Adv Mater 31(42) (2019).
[25] D. Zander, P. Zaslansky, N.A. Zumdick, M. Felten, C. Schnatterer, V.F. Chaineux, J.U. Hammel, M. Storm, F. Wilde, C. Fleck, The Effect of Chemistry and 3D Microstructural Architecture on Corrosion of Biodegradable Mg-Ca-Zn Alloys, Advanced Engineering Materials (2021).
[26] R. Paris, Sur les alliages ternaires magnésium-zinc-calcium, Comptes rendus hebdomadaires des séances de l'académie des sciences 197 (1933) 1634-1636.
[27] R. Paris, Contribution à l'étude des alliages ternaires, Publications scientifiques et techniques du ministère de l'air, Ministère de L'Air 45 (1934) 1-86.
[28] J.B. Clark, The Solid Constitution in the Magnesium-Rich Region of the Mg-Ca-Zn Phase Diagram, T Metall Soc AIME 221(3) (1961) 644-645.
[29] J.B. Clark, JCPDS, Card 12-0266 (1961).
[30] J.C. Slater, Atomic Radii in Crystals, J Chem Phys 41(10) (1964) 3199-&.
[31] T.V. Larionova, W.W. Park, B.S. You, A ternary phase observed in rapidly solidified Mg-Ca-Zn alloys, Scripta Mater 45(1) (2001) 7-12.
[32] P.M. Jardim, G. Solorzano, J.B. Vander Sande, Precipitate crystal structure determination in melt spun Mg-1.5wt%Ca-6wt%Zn alloy, Microsc Microanal 8(6) (2002) 487-496.
[33] P.M. Jardim, G. Solorzano, J.B. Vander Sande, Second phase formation in melt-spun Mg-Ca-Zn alloys, Mat Sci Eng A-Struct 381(1-2) (2004) 196-205.
[34] G. Levi, S. Avraham, A. Zilberov, M. Bamberger, Solidification, solution treatment and age hardening of a Mg-1.6 wt.% Ca-3.2 wt.% Zn alloy, Acta Mater 54(2) (2006) 523-530.
[35] M. Bamberger, G. Levi, J.B.V. Sande, Precipitation hardening in Mg-Ca-Zn alloys, Metall Mater Trans A 37(2) (2006) 481-487.
[36] J.C. Oh, T. Ohkubo, T. Mukai, K. Hono, TEM and 3DAP characterization of an age-hardened Mg-Ca-Zn alloy, Scripta Mater 53(6) (2005) 675-679.
[37] Y.-N. Zhang, D. Kevorkov, J. Li, E. Essadiqi, M. Medraj, Determination of the solubility range and crystal structure of the Mg-rich ternary compound in the Ca-Mg-Zn system, Intermetallics 18(12) (2010) 2404-2411.
[38] Y.-N. Zhang, D. Kevorkov, F. Bridier, M. Medraj, Experimental study of the Ca-Mg-Zn system using diffusion couples and key alloys, Sci Technol Adv Mat 12(2) (2011).
[39] M. Mezbahul-Islam, Y.N. Zhang, C. Shekhar, M. Medraj, Critical assessment and thermodynamic modeling of Mg-Ca-Zn system supported by key experiments, Calphad 46 (2014) 134-147.

[40] H.-X. Li, Y.-P. Ren, Q.-Q. Ma, M. Jiang, G.-W. Qin, Ternary compounds and solid-state phase equilibria in Mg-rich side of Mg-Zn-Ca system at 300 degrees C, T Nonferr Metal Soc 21(10) (2011) 2147-2153.
[41] B. Langelier, X. Wang, S. Esmaeili, Evolution of precipitation during non-isothermal ageing of an Mg-Ca-Zn alloy with high Ca content, Mat Sci Eng A-Struct 538 (2012) 246-251.
[42] www.factsage.com, Version 6.0 (2009)).
[43] K. Kubok, L. Litynska-Dobrzynska, J. Wojewoda-Budka, A. Goral, A. Debski, investigation of structures in as-cast alloys from the Mg-Zn-Ca system, Arch Metall Mater 58(2) (2013) 329-333.
[44] J.D. Cao, T. Weber, R. Schäublin, J.F. Löffler, Equilibrium ternary intermetallic phase in the Mg-Zn-Ca system, J Mater Res 31(14) (2016) 2147-2155.
[45] S. Wasiur-Rahman, M. Medraj, Critical assessment and thermodynamic modeling of the binary Mg-Zn, Ca-Zn and ternary Mg-Ca-Zn systems, Intermetallics 17(10) (2009) 847-864.
[46] M. Cihova, E. Martinelli, P. Schmutz, A. Myrissa, R. Schäublin, A.M. Weinberg, P.J. Uggowitzer, J.F. Loffler, The role of zinc in the biocorrosion behavior of resorbable Mg-Zn-Ca alloys, Acta Biomater 100 (2019) 398-414.
[47] J.F. Löffler, P. Uggowitzer, C. Wegmann, M. Becker, H. Feichtinger, Verfahren und vorrichtung zur vakuumdestillation von hochreinem magnesium, Google Patents, 2014.
[48] K. Urban, Application of High-Voltage Electron-Microscopy to Low-Temperature Radiation-Damage Studies in Metals, J Microsc 97 (1973) 121-127.
[49] M. Klinger, A. Jager, Crystallographic Tool Box (CrysTBox): automated tools for transmission electron microscopists and crystallographers, J Appl Crystallogr 48 (2015) 2012-2018.
[50] P.A. Stadelmann, EMS - A software package for electron diffraction analysis and HREM image simulation in materials science, Ultramicroscopy 21(2) (1987) 131-145.
[51] O.C. Hellman, J.A. Vandenbroucke, J. Rusing, D. Isheim, D.N. Seidman, Analysis of three-dimensional atom-probe data by the proximity histogram, Microsc Microanal 6(5) (2000) 437-444.
[52] A. Cerezo, L. Davin, Aspects of the observation of clusters in the 3-dimensional atom probe, Surf Interface Anal 39(2-3) (2007) 184-188.
[53] D. Vaumousse, A. Cerezo, P.J. Warren, A procedure for quantification of precipitate microstructures from three-dimensional atom probe data, Ultramicroscopy 95(1-4) (2003) 215-221.
[54] A. Nayeb-Hashemi, Phase diagrams of binary magnesium alloys, ASM International, Metals Park, Ohio 44073, USA, 1988.
[55] H. Okamoto, Supplemental Literature Review of Binary Phase Diagrams: Cs-In, Cs-K, Cs-Rb, Eu-In, Ho-Mn, K-Rb, Li-Mg, Mg-Nd, Mg-Zn, Mn-Sm, O-Sb, and Si-Sr, J Phase Equilib Diff 34 (3) (2013) 251-263.
[56] B.C. Zhou, S.L. Shang, Y. Wang, Z.K. Liu, Diffusion coefficients of alloying elements in dilute Mg alloys: A comprehensive first-principles study, Acta Mater 103 (2016) 573-586.
[57] W.D. Callister, Materials Science and Engineering: An Introduction, New York: Wiley 2003
[58] I. Basu, M. Chen, J. Wheeler, R.E. Schäublin, J.F. Loffler, Stacking-fault mediated plasticity and strengthening in lean, rare-earth free magnesium alloys, Acta Mater 211 (2021).
[59] M.K. Miller, M.G. Hetherington, Local Magnification Effects in the Atom Probe, Surf Sci 246 (1-3) (1991) 442-449.
[60] M. Cihova, R. Schäublin, L.B. Hauser, S.S.A. Gerstl, C. Simson, P.J. Uggowitzer, J.F. Loffler, Rational design of a lean magnesium-based alloy with high age-hardening response, Acta Mater 158 (2018) 214-229.
[61] G.K.H. Kolb, S. Scheiber, H. Antrekowitsch, P.J. Uggowitzer, D. Poschmann, S. Pogatscher, Differential Scanning Calorimetry and Thermodynamic Predictions-A Comparative Study of Al-Zn-Mg-Cu Alloys, Metals-Basel 6 (8) (2016) 180.
[62] A. Deschamps, Y. Brechet, Influence of predeformation and ageing of an Al–Zn–Mg alloy—II. Modeling of precipitation kinetics and yield stress, Acta Mater 47(1) (1998) 293-305.
[63] W.F. Hosford, Mechanical Behavior of Materials, Cambridge University Press, 2010.
[64] J. Friedel, Dislocations, Oxford, 1964.

**Supplementary Information**

**Precipitation in lean Mg–Zn–Ca alloys**

*R. E. Schäublin\*, M. Becker, M. Cihova, S. S. A. Gerstl, D. Deiana, C. Hébert, S. Pogatscher, P. J. Uggowitzer, and J. F. Löffler*

The crystallographic structure of ternary Ca–Mg–Zn precipitates in ZX20 alloys extruded at 330 °C has been scrutinized in detail in order to remove any ambiguities on its identity. Figure S1 presents the three candidate crystallographic structures identified in the literature. One notes in these [0001] views the clear visual difference between $Ca_2Mg_6Zn_3$ on the one hand, and $Ca_2Mg_5Zn_5$ and $Ca_3Mg_{11}Zn_4$ on the other, which share the same symmetry. The latter two structures exhibit a periodic hexagonal pattern that is not present in $Ca_2Mg_6Zn_3$.

Figures S2 and S3 display an analysis made on the ultrahigh-resolution HAADF STEM images and chemical maps taken along [0001] and [$2\bar{1}\bar{1}0$], respectively (presented in Fig. 11). The analysis is based on a 2D projection of the candidate crystallographic structures, convoluted with a Gaussian to represent the electron beam profile. The projection is made either with all atoms, representing the HAADF STEM images, or with only Ca, Mg or Zn atoms to represent the respective elemental maps. It appears that the HAADF STEM image analysis in both directions (Figs. S2 and S3) clearly favors $Ca_2Mg_5Zn_5$ as the best candidate. The chemical maps, however, still generate some ambiguities. In the [0001] direction (Fig. S2) the Zn map presents the experimentally observed periodic hexagonal contrast, pointing again at $Ca_2Mg_5Zn_5$ as the best candidate, but the Mg map may favor $Ca_3Mg_{11}Zn_4$. In the [$2\bar{1}\bar{1}0$] direction (Fig. S3), the Zn map presents wavy horizontal lines that seem to be present in the experimental map, again pointing at $Ca_2Mg_5Zn_5$, while the other maps do not allow a conclusion. This analysis reveals that while a simple structure projection may be helpful to interpret a STEM image or a chemical map, one should note that it does not include the real electron-matter interactions, which may be at the origin of this ambiguity.

In order to come to a firm conclusion, we thus simulated the HAADF STEM images using the parameters of the experimental acquisition. In a first step, three simulation software were compared to make sure that the result does not depend on the chosen algorithm. As a test, the HAADF STEM image of $Ca_2Mg_5Zn_5$ in the [0001] direction was considered and compared to an experimental image (Fig. S4). It appears that all software yield the same contrast, with a periodic hexagonal pattern, a similar contrast evolution with increasing thickness, and a qualitatively good match to the experimental image (Fig. S4). Dr. Probe software was selected for further simulations.

In general, we observed contrast variations across the HAADF STEM images in the ternary Ca–Mg–Zn precipitates. Among others, one can attribute them to a structure variation across the precipitate, for example with a change of the Zn-to-Mg ratio that may encompass both $Ca_2Mg_5Zn_5$ and $Ca_3Mg_{11}Zn_4$.

In order to raise this doubt, a typical HAADF STEM image taken in the [0001] direction was analyzed with the help of simulations. The results are presented in Fig. S5. The image is best matched with $Ca_2Mg_5Zn_5$, but not with $Ca_3Mg_{11}Zn_4$, because the hexagonal pattern cannot be met with the latter. The visible contrast variations across the image of Fig. S5 can also be clearly attributed to a thickness change, which allowed narrowing down the sample thickness to 40 – 70 nm.

After these successful tests, HAADF STEM images of the ternary Ca–Mg–Zn precipitates were taken in the [0001] and [$2\bar{1}\bar{1}0$] directions for ZX20 (presented in Fig. 12) and in the [$10\bar{1}0$] direction for ZX10 (another extruded Mg–Zn–Ca alloy: Mg-1Zn-0.3Ca, in wt.% [1]) and analyzed by simulations (Fig. S6). All images are best matched with $Ca_2Mg_5Zn_5$, but not with $Ca_3Mg_{11}Zn_4$. We thus conclude that the structure of the ternary Ca–Mg–Zn precipitates in extruded lean Mg–Zn–Ca alloys is $Ca_2Mg_5Zn_5$. In addition, one can also state that HAADF STEM imaging combined with supporting simulations is generally sufficient to resolve crystallographic structures.

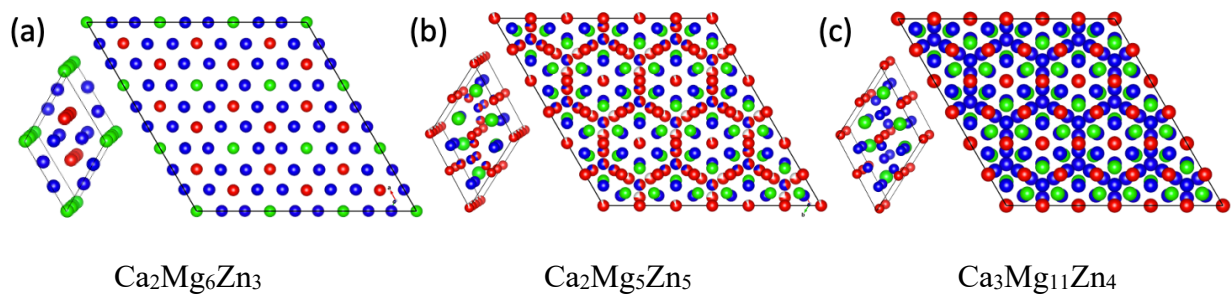

        $Ca_2Mg_6Zn_3$          $Ca_2Mg_5Zn_5$          $Ca_3Mg_{11}Zn_4$

Figure S1. Candidate structures for the Ca–Mg–Zn ternary phases reported for ZX20 heat-treated at 350 °C. (a) $Ca_2Mg_6Zn_3$, (b) $Ca_2Mg_5Zn_5$ and (c) $Ca_3Mg_{11}Zn_4$. The color code is green for Ca, blue for Mg, and red for Zn. The unit cell is displayed in perspective view and a portion of the crystal made of 3×3×2 unit cells is viewed along [0001]. Among the three structures, $Ca_2Mg_6Zn_3$ sticks out, while $Ca_2Mg_5Zn_5$ and $Ca_3Mg_{11}Zn_4$ are similar, sharing the same symmetry ($P6_3/mmc$) and revealing in the images a similar dense arrangement of atoms with a hexagonal pattern. The main difference between $Ca_2Mg_5Zn_5$ and $Ca_3Mg_{11}Zn_4$ is their Zn-to-Mg ratio, which is seen in the hexagons' outline, with $Ca_2Mg_5Zn_5$ presenting more Zn atoms than Mg.

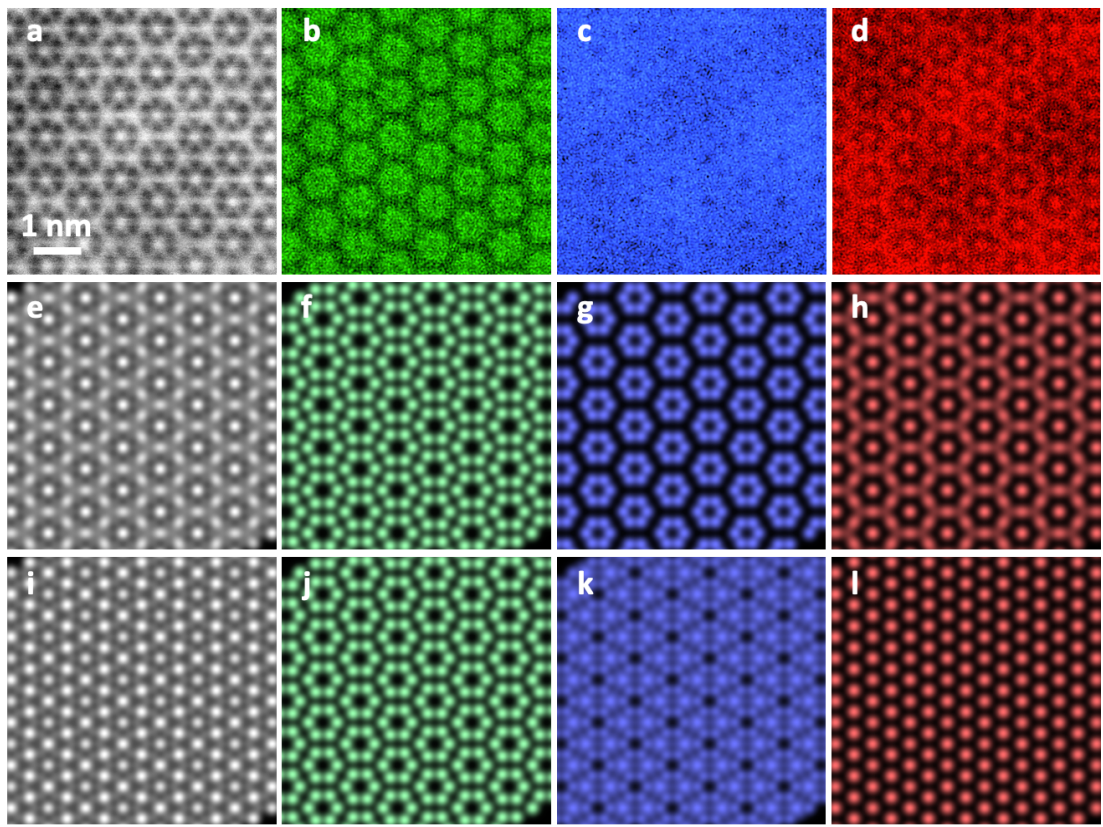

Figure S2. (a) Experimental HAADF STEM images and corresponding EDS chemical maps of (b) Ca in green, (c) Mg in blue and (d) Zn in red, of a ternary precipitate in ZX20 extruded at 330 °C viewed along [0001] (presented in Fig. 11). The images were modeled by projecting the atom positions of the candidate crystal structures along [0001]. The projections are convoluted with a Gaussian to represent the electron beam with a full width at half maximum (FWHM) of 1 Å. The resulting projections are presented in (e-h) for $Ca_2Mg_5Zn_5$ and (i-l) for $Ca_3Mg_{11}Zn_4$. It appears that the best match for (a) the HAADF STEM image is obtained with (e) $Ca_2Mg_5Zn_5$, revealing a hexagonal pattern. The Ca map (b) does not allow to distinguish between the two structures because both projections (f, j) yield the same contrasts, with a hexagonal pattern similar to that in the experimental map. The Mg map (c) presents a faint periodic lattice of dark spots over a uniform background. The projection of either structure (g, k) somewhat matches the periodic lattice, but $Ca_3Mg_{11}Zn_4$ (k) yields a better match to the background. The Zn map (d) exhibits a clear pattern of hexagons with a bright spot in their center. The projected map that provides the best match is obtained with $Ca_2Mg_5Zn_5$ (h).

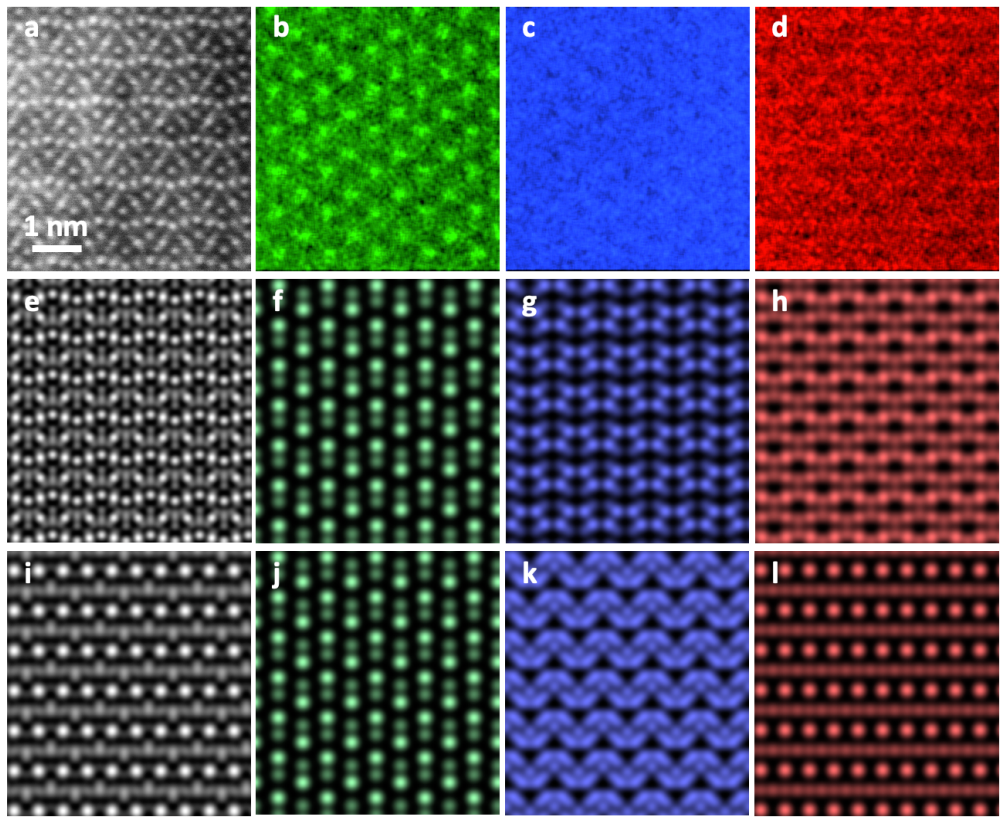

Figure S3. (a) Experimental HAADF STEM images and corresponding EDS chemical maps of (b) Ca in green, (c) Mg in blue and (d) Zn in red, of a ternary precipitate in ZX20 extruded at 330 °C viewed along $[2\bar{1}\bar{1}0]$ (presented in Fig. 11). The images were modeled by projecting the atoms position of the candidate crystal structures along $[2\bar{1}\bar{1}0]$. The projections are convoluted with a Gaussian to represent the electron beam with a FWHM of 1 Å. The resulting projections are presented in (e-h) for $Ca_2Mg_5Zn_5$ and in (i-l) for $Ca_3Mg_{11}Zn_4$. It appears that the best match for (a) the HAADF STEM image is obtained with (e) $Ca_2Mg_5Zn_5$. The Ca map (b) does not allow to distinguish between the two structures because both projections (f, j) yield the same contrast, with an appearance similar to that of the experimental map. The Mg map (c) presents a faint periodic lattice of dark contrasts over a uniform background. The projections (g, k) hardly match the experimental structure. The Zn map (d) exhibits a faint horizontal wavy-lines contrast over a uniform background. The projected map that comes closer to it is obtained with $Ca_2Mg_5Zn_5$ (h).

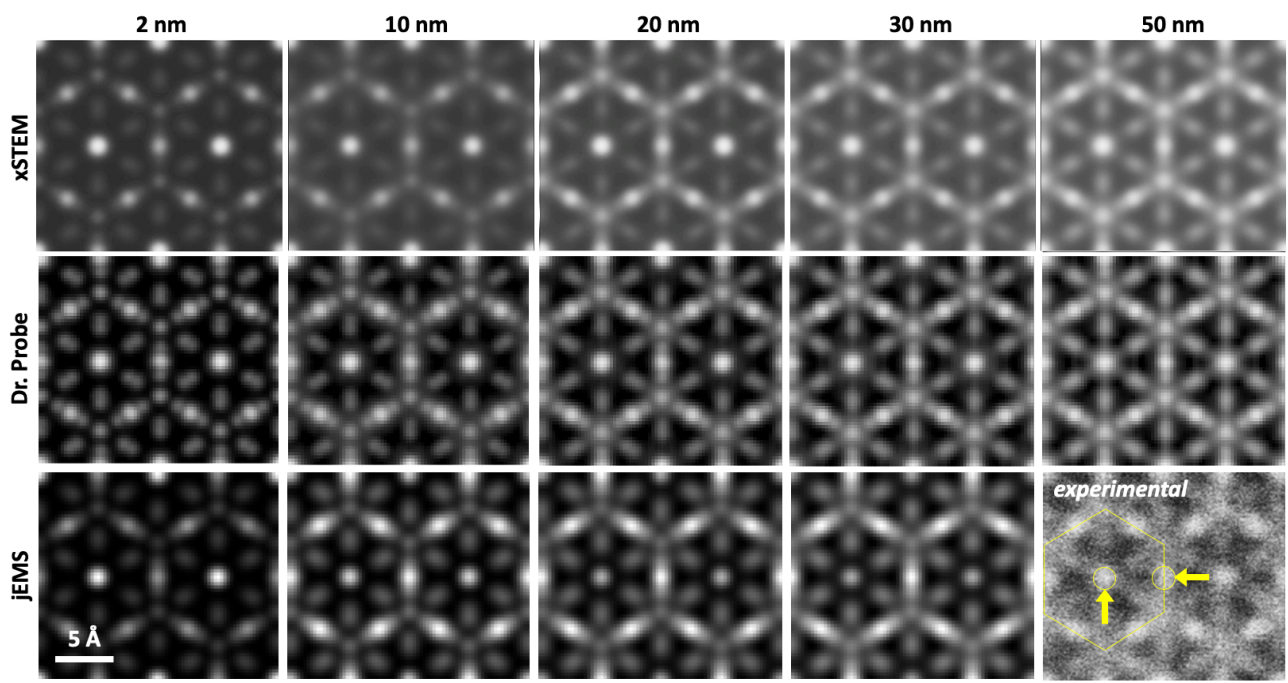

Figure S4. Comparison between STEM image simulation software, namely (first row) xSTEM [1], (middle row) Dr. Probe [2] and (bottom row) jEMS [3] for various sample thicknesses (as indicated at the top). The experimental image (bottom right) used as a reference for this test is an ultrahigh-resolution HAADF STEM micrograph taken along [0001] in a Ca–Mg–Zn ternary precipitate present in ZX20 extruded at 330 °C. All software yield the same type of contrast, exhibiting a clear pattern of hexagons with a bright spot in their center (vertical arrow). The hexagons' outline is increasingly discernable with increasing sample thickness. The central spot intensity remains relatively constant (xSTEM, Dr. Probe) or decreases (jEMS) with increasing sample thickness, while the bright contrast in the center of each hexagons' edge (horizontal arrow) gets dimmer relative to the hexagons' outline.

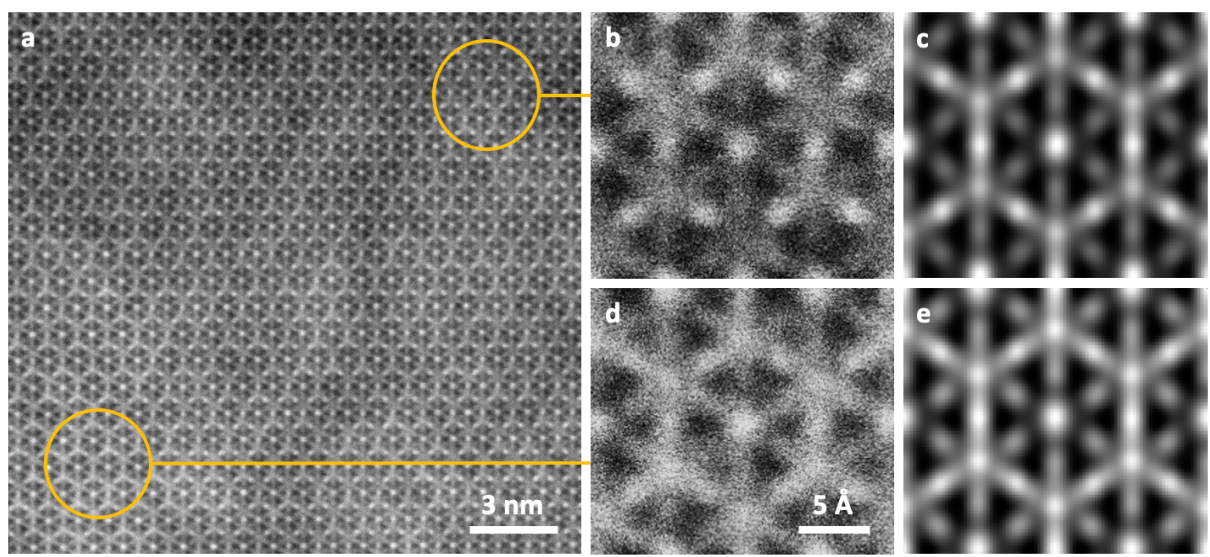

Figure S5. (a) Typical ultrahigh-resolution HAADF STEM micrograph taken along [0001] in Ca–Mg–Zn ternary precipitates of ZX20 extruded at 330 °C, exhibiting a periodic pattern made of hexagons whose center presents a bright and relatively constant contrast. There are, however, contrast variations across the image, ranging from marked bright spots in the center of the hexagons' edges (b) to marked hexagons (d). The images simulated with Dr. Probe software [2], which best match the images (b) and (d), are obtained with the $Ca_2Mg_5Zn_5$ structure and for a sample thickness of (c) 40 nm and (e) 70 nm, respectively. This illustrates that the contrast variations are due to thickness variations and not to changes in structure.

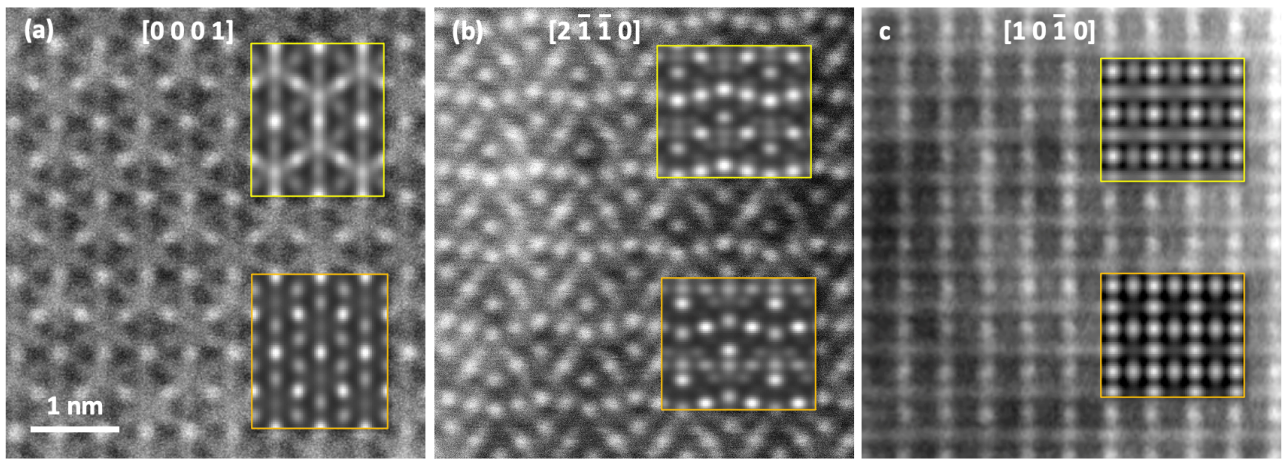

Figure S6. Typical ultrahigh-resolution HAADF STEM micrographs taken along (a) [0001] and (b) [2$\bar{1}\bar{1}$0] in Ca–Mg–Zn ternary precipitates present in a ZX20 alloy extruded at 330 °C. The micrograph shown in (c) was taken along [10$\bar{1}$0] in a Ca–Mg–Zn ternary precipitate present in ZX10 (Mg–Zn1.0–Ca0.3, in wt.% [4]) extruded at 300 °C (from Fig. S3e in [5]). The inserts present the best matching images simulated with Dr. Probe software [2] using either (top inserts) $Ca_2Mg_5Zn_5$ or (bottom inserts) $Ca_3Mg_{11}Zn_4$. With the $Ca_2Mg_5Zn_5$ structure (top inserts), there is an unequivocal good match to the three images. One notes in particular in (a) the bright hexagonal outlines, in (b) the horizontal wavy spots line of similar brightness and in (c) the horizontal and straight bright lines, all being unmatched with $Ca_3Mg_{11}Zn_4$.

**References Supplementary Information**

[1] K. Ishizuka, A practical approach for STEM image simulation based on the FFT multislice method, Ultramicroscopy 90(2-3) (2002) 71-83.
[2] J. Barthel, Dr. Probe: A software for high-resolution STEM image simulation, Ultramicroscopy 193 (2018) 1-11.
[3] P.A. Stadelmann, EMS - A software package for electron diffraction analysis and HREM image simulation in materials science, Ultramicroscopy 21(2) (1987) 131-145.
[4] J. Hofstetter, M. Becker, E. Martinelli, A.M. Weinberg, B. Mingler, H. Kilian, S. Pogatscher, P.J. Uggowitzer, J.F. Loffler, High-Strength Low-Alloy (HSLA) Mg-Zn-Ca Alloys with Excellent Biodegradation Performance, JOM 66(4) (2014) 566-572.
[5] M. Cihova, P. Schmutz, R. Schäublin, J.F. Loffler, Biocorrosion Zoomed In: Evidence for Dealloying of Nanometric Intermetallic Particles in Magnesium Alloys, Adv Mater 31(42) (2019) .